\documentclass[]{aa}

\usepackage{natbib}
\bibpunct{(}{)}{;}{a}{}{,} 
\usepackage{graphicx}
\usepackage{txfonts}
\usepackage[]{hyperref}
\begin{document}

   \title{\textit{Gaia} Data Release 2: Validation of the classification of RR Lyrae and Cepheid variables with the \textit{Kepler} and \textit{K2} missions}
   \titlerunning{DR2: Validation of RR Lyraes and Cepheids with \textit{Kepler} and \textit{K2}}
   \author{L\'aszl\'o Moln\'ar\inst{1,2}
          \and
          Emese Plachy\inst{1,2}
          \and
          \'Aron L. Juh\'asz\inst{1,3}
          \and
          Lorenzo Rimoldini\inst{4}
          }

   \institute{Konkoly Observatory, MTA CSFK, H-1121, Budapest, Konkoly Thege Mikl\'os \'ut 15-17., Hungary
         \and
             MTA CSFK Lend\"ulet Near-Field Cosmology Research Group, H-1121, Budapest, Konkoly Thege Mikl\'os \'ut 15-17., Hungary
         \and
             Department of Astronomy, E\"otv\"os Lor\'and University, H-1117, Budapest, P\'azm\'any P\'eter s\'et\'any 1/a, Hungary
          \and 
             Department of Astronomy, University of Geneva, Chemin d'Ecogia 16, CH-1290, Versoix, Switzerland}

   \date{Received ... / accepted ...}

 \abstract
{The second data release of the \textit{Gaia} mission includes an advance catalog of variable stars. The classification of these stars are based on sparse photometry from the first 22 months of the mission.}
{We set out to investigate the purity and completeness of the all-sky \textit{Gaia} classification results with the help of the continuous light curves of the observed targets from the \textit{Kepler} and \textit{K2} missions, focusing specifically on RR~Lyrae and Cepheid pulsators, outside the Galactic Bulge region.}
{We crossmatched the \textit{Gaia} identifications with the observations collected by the \textit{Kepler} space telescope. We inspected the light curves visually, then calculated the relative Fourier coefficients and period ratios for the single- and double-mode \textit{K2} RR Lyrae stars to further classify them.}
{We identified 1443 and 41 stars classified as RR Lyrae or Cepheid variables in \textit{Gaia} DR2 in the targeted observations of the two missions and 263 more RR Lyre targets in the Full-Frame Images (FFI) of the original mission. We provide the crossmatch of these sources. We conclude that the RR~Lyrae catalog has a completeness between 70--78\%, and provide a purity estimate between 92--98\% (targeted observations) with lower limits of 75\% (FFI stars) and 51\% (\textit{K2} worst-case scenario). The low number of Cepheids prevents us from drawing detailed conclusions but the purity of the DR2 sample is estimated to be around 66\%. }
{}

   \maketitle
%
\section{Introduction}
RR Lyrae stars are large-amplitude pulsating stars with easily recognizable light curve shapes. They can be used to measure distances, to trace structures in the Milky Way and other galaxies, and to investigate stellar pulsation and evolution. Large sky surveys have mapped the distribution of RR Lyrae stars in particular to large numbers and depths \citep[see, e.g.,][for some examples]{drake2013,sesar2013,beaton2016,hernitschek2016,minniti2017}. 

Cepheids, an umbrella term used to describe the classical $\delta$ Cephei stars, the Type II Cepheids, and the anomalous Cepheids, are even more important tools to map the structure of the cosmos and individual galaxies: although they are less frequent than RR Lyrae stars, they are more luminous, and hence more easily detectable. The OGLE survey had very thoroughly searched both the Galactic bulge and the Magellanic Clouds for all members of the Cepheid and RR Lyrae families. According to \citet{soszynksi2017}, their collection of classical Cepheids is now near-complete, concluding the work started a century ago by \citet{leavitt1908}.

The \textit{Gaia} Data Release 2 \citep[DR2;][]{brown2018} represents a new step forward in mapping the distribution of RR Lyrae and Cepheid stars, as well as other variable stars in the Milky Way and its vicinity. However, these classifications are based on sparse photometry, and those classifications can be affected by poor phase coverage and/or confusion between variable types.

Space photometric missions, such as \textit{MOST}, \textit{CoRoT}, \textit{Kepler} or \textit{BRITE}, provide dense, continuous photometry of variable stars that could be, in principle, used to validate the results from other surveys. However, most missions have either small apertures or very limited sample sizes, or both, therefore their use for validation purposes is very limited. The only exception is the \textit{Kepler} space telescope that observed hundreds of thousands of targets so far, including thousands of RR Lyrae and hundreds of Cepheid stars \citep{szabo2017,molnar2018}. Our preliminary studies with the PanSTARRS PS1 survey by \citet{hernitschek2016}, indicate that a comparison with data from \textit{Kepler} is feasible, however, the first results indicate that eclipsing binary stars may contaminate the RR Lyrae sample in the Galactic disk in that survey \citep{juhasz2018}. 

These findings led us to use the observations of \textit{Kepler} to provide an independent validation of the classifications of the various RR Lyrae and Cepheid-type variables in \textit{Gaia} DR2. Conversely, input from \textit{Gaia} DR2 will allow us in the future to correctly identify targets as RR Lyrae or Cepheid stars hiding in the \textit{Kepler} and K2 databases.

The paper is structured as follows: in Sect.~\ref{sect:obs} we introduce the variable star observations made by \textit{Gaia} and \textit{Kepler}, respectively; in Sect.~\ref{sect:id} we describe the classification of the \textit{Kepler} light curves into the various subclasses; in Sect.~\ref{sect:res} we present the results of the validation, and in Sect.~\ref{sect:concl} we provide our conclusions.

\section{Observations}
\label{sect:obs}

\subsection{Variable stars in \textit{Gaia} DR2}

\citet{holl2018} summarize the results regarding the over half a million variable stars published in \textit{Gaia} DR2, which include candidates of RR~Lyrae stars, Cepheids, long period variables, rotation modulation (BY~Draconis) stars, $\delta$~Scuti \& SX~Phoenicis stars, and short-timescale variables.

Different techniques were employed in the variability pipeline depending on the variability type. Herein, we focus on the Cepheids and RR~Lyrae stars identified by the all-sky classification (Rimoldini et al., in preparation), available in the \texttt{gaiadr2.vari\_classifier\_result} table of the \textit{Gaia} archive. Such classifications are normally verified by a subsequent pipeline module dedicated to the RR~Lyrae and Cepheid variables \citep{Clementini2018}, but only for sources with at least 12~field-of-view (FoV) transits in the \textit{G} band. The variable classes that we use are the four RR~Lyrae subtypes, RRAB, RRC, and RRD/ARRD (corresponding to fundamental-mode, first-overtone, and double-mode stars with either normal or anomalous period ratios) and various Cepheid types, CEP, T2CEP, and ACEP (corresponding to classical or $\delta$ Cephei stars, Type II Cepheids, and anomalous Cepheids). Throughout the paper, we refer to the variable classifications in \textit{Gaia} DR2 in all capitals (RRAB, RRC, RRD), and use the regular RRab, RRc, RRd notations for our classifications based on the \textit{Kepler} and \textit{K2} light curves.

Our goal is to provide an independent assessment of the \textit{completeness} and \textit{purity} rates of the all-sky classification results of  RR~Lyrae and Cepheid candidates based on the fine light-curve sampling of \textit{Kepler} and \textit{K2} targets, regardless of the number of observations and other features or limitations related to the \textit{Gaia} data. We define completeness here as the fraction of sources with properly identified classes in \textit{Gaia} DR2 against all RR Lyrae stars observed in the \textit{Kepler} and K2 missions. Purity is defined here as the fraction of sources with properly identified variable classes compared to all sources in those classes in \textit{Gaia} DR2.

\textit{Kepler} is superior in terms of photometric precision and duty cycle to all ground--based surveys. The OGLE survey comes close due to the decade-long coverage, but it is limited to the Galactic Bulge, disk, and the Magellanic Clouds \citep{ogleiv}, whereas \textit{Kepler} observed several halo fields. We note that \citet{holl2018} presents some completeness and purity estimates based on the OGLE survey results for the RR Lyrae and Cepheid candidates. The only other program similar to \textit{Kepler} is the Transiting Exoplanet Survey Satellite space telescope \citep[\textit{TESS},][]{tess}: it has already started collecting continuous photometry from nearly the whole sky, but it will not reach the depth of either \textit{Gaia} or  \textit{Kepler} or ground-based surveys.

In this study, we used the following information available in the \textit{Gaia} DR2 archive (fields and table names specified in footnotes): celestial coordinates,\footnote{\texttt{\tiny ra} and \texttt{\tiny dec} from \texttt{\tiny gaiadr2.gaia\_source}} median \textit{G} brightnesses,\footnote{\texttt{\tiny median\_mag\_g\_fov} from \texttt{\tiny gaiadr2.vari\_time\_series\_statistics}} 
the number of field-of-view transits in the \textit{G}~band,\footnote{\texttt{\tiny num\_selected\_g\_fov} from \texttt{\tiny gaiadr2.vari\_time\_series\_statistics}} classification classes\footnote{\texttt{\tiny best\_class\_name} from \texttt{\tiny gaiadr2.vari\_classifier\_result}} and scores\footnote{\texttt{\tiny best\_class\_score} from \texttt{\tiny gaiadr2.vari\_classifier\_result}} ranging between 0–1, provided by the all-sky classification pipeline (Rimoldini et al., in preparation).


\subsection{The \textit{Kepler} and \textit{K2} missions}

The \textit{Kepler} space telescope was designed to detect transiting exoplanets and determine the occurrence of Earth-like, temperate rocky planets around Sun-like stars \citep{borucki2010,Borucki-2016}. To achieve that, it collected quasi-continuous photometry of approximately $170\,000$ stars in a field between Lyra and Cygnus, spanning about 115~deg$^2$. The majority of stars were observed with a 29.4~min sampling, called long cadence (LC), and the telescope was rolled by 90~deg every quarter year. Not all stars were continuously observed: the target list was updated for every quarter. The prime mission lasted for 4~years, until the breakdown of two reaction wheels on board. 

Afterwards, \textit{Kepler} was reoriented to observe in shorter, 60-80~day long campaigns along the Ecliptic, using only the two remaining reaction wheels. The new mission, named \textit{K2}, had no core science program, and the space telescope has been utilized as a space photometric observatory instead, carrying out a very broad science program that extends beyond exoplanets and stellar physics \citep{howell-k2}. Coordination and target selection is managed by the Kepler Guest Observer Office\footnote{http://keplerscience.arc.nasa.gov}. Stellar physics studies are coordinated by the \textit{Kepler} Asteroseismic Science Consortium (KASC), which has a dedicated RR Lyraes and Cepheids Working Group.

During the original mission, \textit{Kepler} observed about 50 RR~Lyrae stars, one classical Cepheid, V1154~Cyg, and a Type-II Cepheid of the RVb subtype, DF Cyg \citep[see, e.g.][and references therein]{benko2010,nemec2013,derekas2017,bodi2016,vega2017}. The \textit{K2} mission greatly expanded the spatial coverage of the telescope, and propelled the numbers of observed RR~Lyrae and Cepheid stars into several thousands and hundreds, respectively, greatly exceeding the observations of previous space photometric missions \citep{szabo2017}.

   \begin{figure}
   \centering
   \includegraphics[width=1.0\columnwidth]{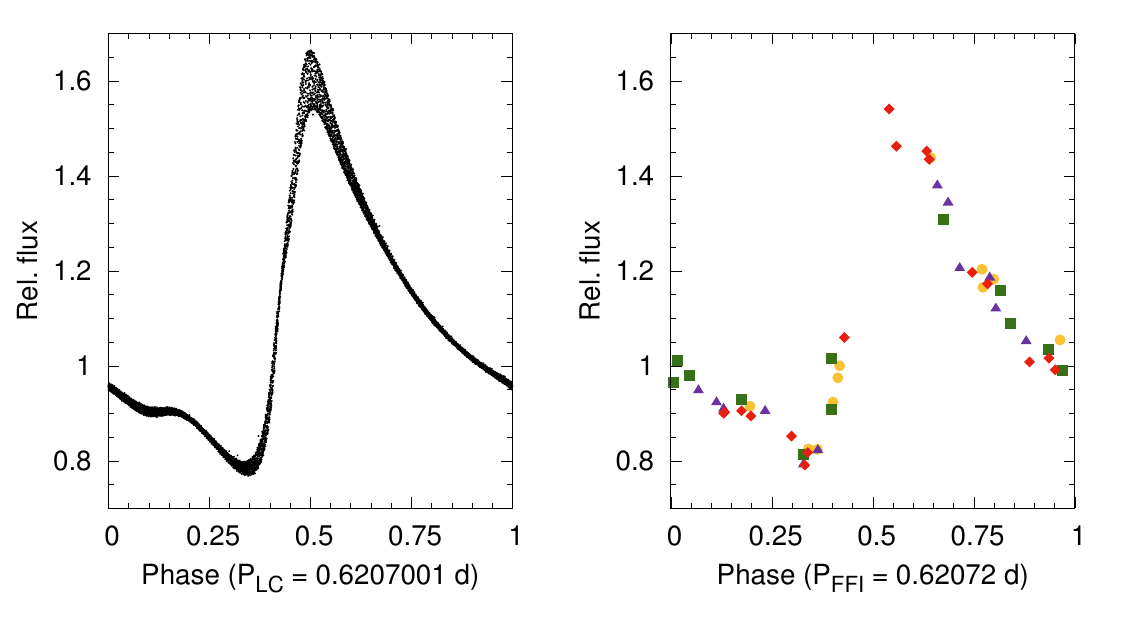}
   \caption{Left: phased long cadence light curve of a modulated RRab star, KIC~5559631, from the original \textit{Kepler} mission, using the tailor-made light curve of \citet{benko2014}. Right: the light curve from the full-frame image of the same star, extracted by the \texttt{f3} code \citep{f3}. Pulsation periods were determined for the two data sets independently. Different colours and symbols denote data from different CCD modules.}
   \label{fig:ffi}
    \end{figure}
    
   \begin{figure*}
   \centering
   \includegraphics[width=1.0\textwidth]{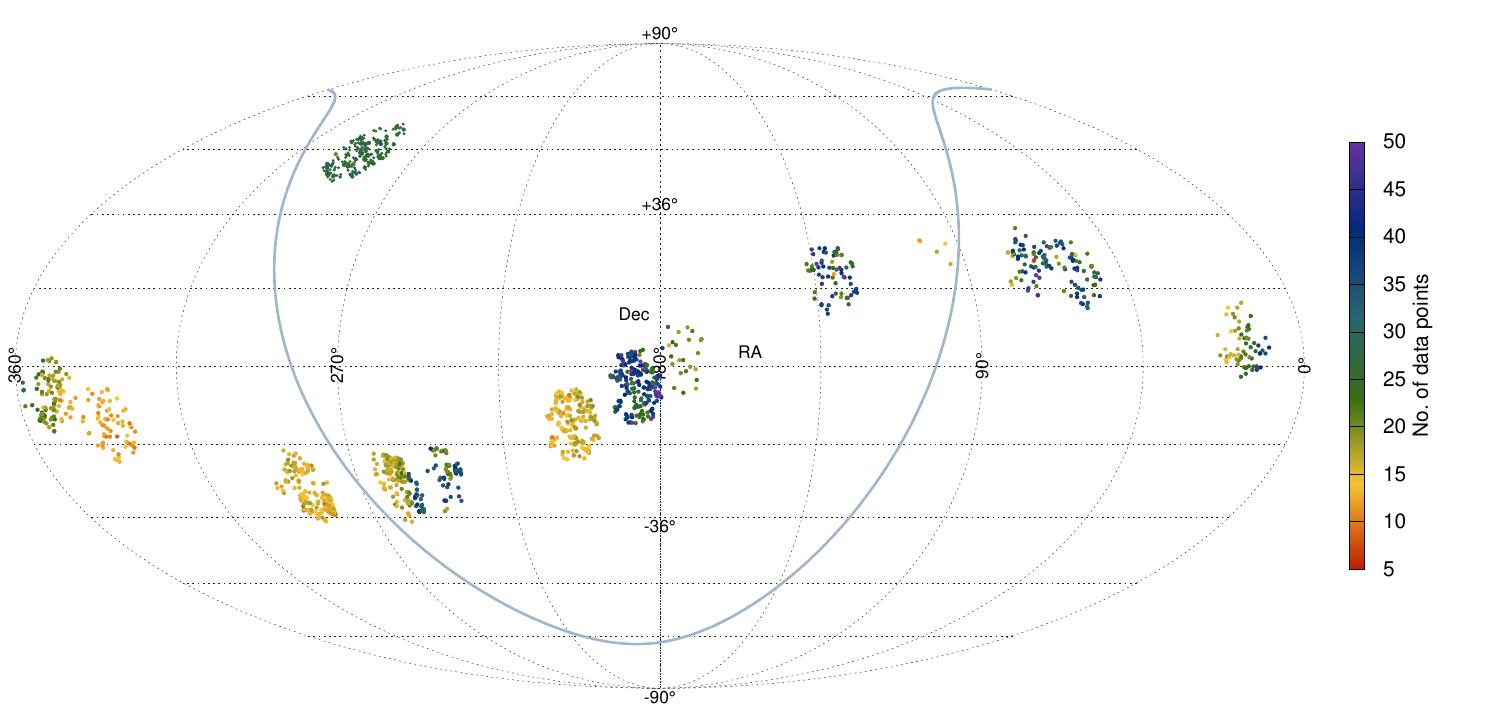}
   \caption{The distribution of the \textit{Gaia} DR2 RR Lyrae candidates confirmed by the \textit{Kepler} (targeted and FFI) and \textit{K2} measurements in the sky, mapped by the Mollweide projection of equatorial coordinates. The thick line in light blue marks the Galactic equator and the colour coding shows the number of \textit{Gaia} field-of-view transits in the $G$~band per star. The original \textit{Kepler} field is the northernmost group of stars at RA~$\approx 300$~deg. }
   \label{fig:rr_map}
    \end{figure*}

An important aspect of the \textit{Kepler} space telescope that sets it apart from other space photometric missions, is the large optical aperture (95~cm), allowing us to reach targets fainter than \textit{Kp} $\sim 20$ mag (where \textit{Kp} refers to brightness in the wide passband of \textit{Kepler}). In case of RR~Lyrae stars, this translates to the distance of the nearest dwarf galaxies \citep{molnar2015}. \textit{Kepler} is the only space telescope that is able to deliver continuous time series photometry while matching the depth of \textit{Gaia} DR2. 
Ground-based, deep, but sparse surveys such as Catalina \citep{2009ApJ...696..870D} and PanSTARRS PS1 \citep[e.g.,][]{2016arXiv161205560C} can also be useful to identify faint RR Lyrae candidates and have been used to select targets for the \textit{K2} mission. But these surveys lack the potential of continuous light curves to unambiguously differentiate pulsating stars from other sources, like eclipsing binaries that could be confused with them in sparse data sets \citep{juhasz2018}. We exploit the properties of the \textit{Kepler} sample mentioned above---depth, coverage, and field-of-view size---to verify and validate the classification of RR~Lyrae and Cepheid variables in \textit{Gaia} DR2.

It is important to note that only pre-selected targets have been observed in both \textit{Kepler} and \textit{K2} missions. Data storage and download bandwidth limitations meant that only a low percentage of pixels were used to gather data in any given observing quarter or campaign. Full-frame images (FFI) were rarely stored, but a sparse sample is available for the original mission.\footnote{52 FFIs were recorded in the original \textit{Kepler} mission, and one or two per campaign in the \textit{K2} mission.} 
In the \textit{K2} mission, targets were exclusively proposed through the Guest Observer program. RR Lyrae candidates from large sky-surveys were revised and the weakest candidates were not proposed \citep{plachy2016}. The target list was then cut by the Guest Observer Office, especially in the early campaigns. Therefore the observed sample is neither a complete sample of the potential RR Lyrae stars nor fully representative of the populations of stars within a field of view. However, these data still represent the largest sample of space-based photometry available to us. At the time of this study, data from Campaigns 0 to 13 were processed and released.

The resolution of \textit{Kepler} at 4"/px is much poorer than that of \textit{Gaia}. This could potentially lead to source confusion. However, RR~Lyrae and Cepheid light curves are easily recognizable even if a target is blended with another star nearby. Therefore, if photometric variation was detected at the target coordinates, we accepted it as a confirmed variable. Multiple stars of the same variable type could still be confused but we found no such examples. Given the limited angular resolution, we decided to avoid the Galactic Bulge areas. The OGLE survey provides superior coverage for classification and validation purposes in the Bulge, and was already utilited in this manner by \citet{holl2018}. The K2 Campaigns 9 and 11 (C9 and C11) targeted the Galactic Bulge: we omitted C9 entirely, and only used stars on C11 below --6~deg Galactic latitude, i.e., excluding the region covered by the OGLE survey. Only a small fraction of the OGLE RR~Lyrae targets were included in the \textit{K2} observations in C11, and their inclusion would not have been representative for the Bulge population. We note that even with these cuts, blending and confusion was hindering the detection and classification of some targets in C11 and C7, the latter which targeted the Sagittarius (Sgr) stream and the outskirts of the Bulge. 

\section{Target identification and classification}
\label{sect:id}
We crossmatched the \textit{Gaia} DR2 sources  that were classified as RR~Lyrae or Cepheid variables around the original \textit{Kepler} Lyra-Cygnus field with the \textit{Kepler} Input Catalog \citep[KIC,][]{kic}. The results of the crossmatch are presented in Appendix~\ref{app}. We did not use the DR2 parallax and $G_{\mathrm{BP}}-G_{\mathrm{RP}}$ colour data for classification and relied only on the light curve shape information provided by \textit{Kepler}. For the original mission, we searched for observed targets and classifications within the literature \citep{benko2010,nemec2013,benko2014,moskalik2015} or visually inspected the light curves available at MAST\footnote{Mikulski Archive for Space Telescopes, \url{http://archive.stsci.edu}}. 

During the original \textit{Kepler} mission, 52~FFIs were recorded: 8 during commissioning, and a further 44 during the mission, each taken before the monthly data downlink period. The integration time was the same as for LC exposures. Although these images only provide sparse photometry, they were obtained at completely different epochs with respect to the \textit{Gaia} observations, and are numerous enough to provide acceptable phase coverage, and a more representative stellar sample that is not affected by selection bias. We used the \texttt{f3} code developed by \citet{f3} to extract the photometry of the stars, then folded the light curves with the most likely period based on the FFI data. A comparison of LC and FFI light curves of an RRab star is illustrated in Fig.~\ref{fig:ffi}, which also shows the agreement of the periods recovered from the two light curves. We compared the FFI periods to the periods determined from normal light curves for all RR~Lyrae-type stars that \textit{Kepler} observed and they agree in all cases. The low number of data points in FFI light curves effectively reduces the faint limit of this data set to $G\sim20$ mag.

   \begin{figure*}
   \centering
   \includegraphics[width=1.0\textwidth]{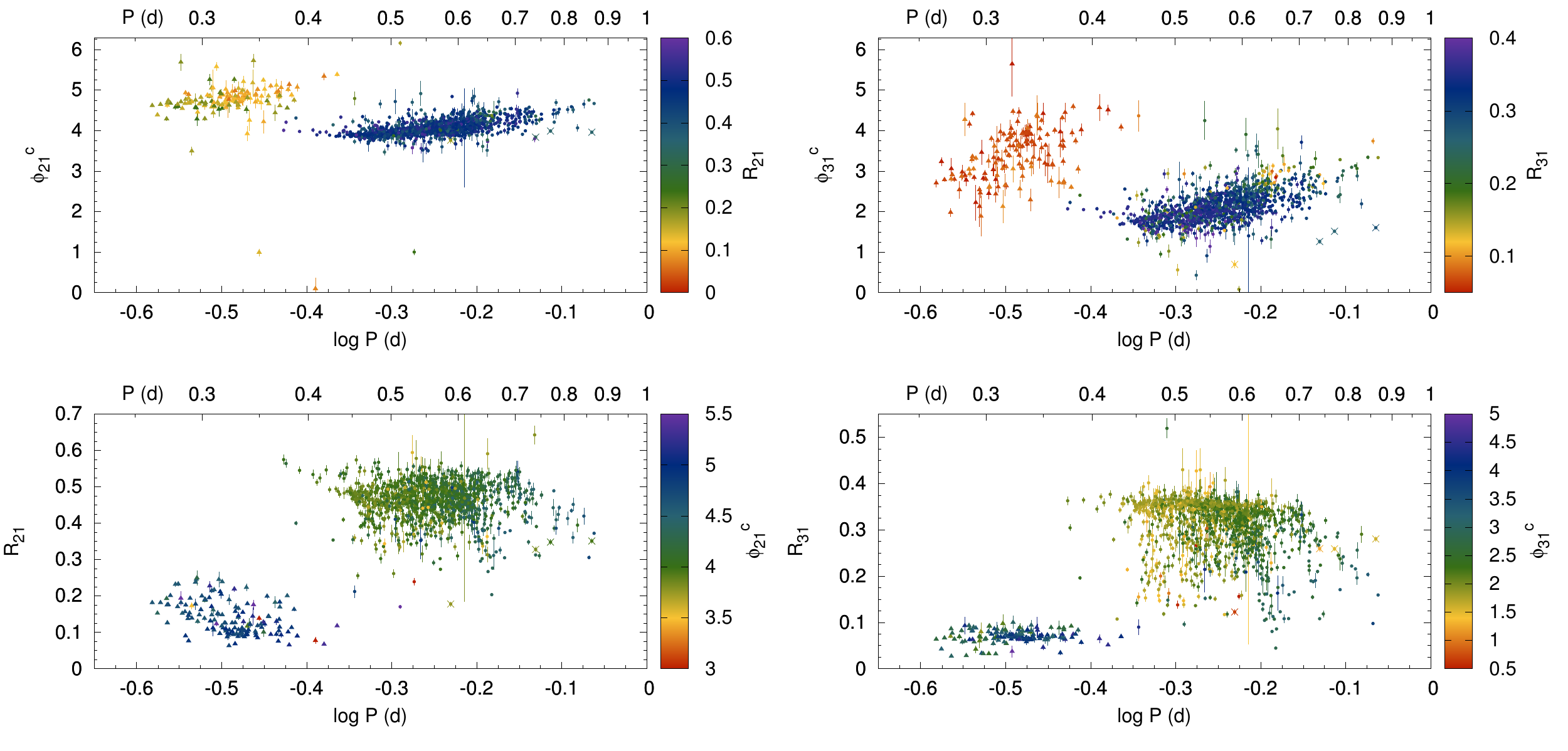}
   \caption{Fourier parameters of the light curves of crossmatched \textit{K2--Gaia} RR~Lyrae-type sources. The $i$th and the first Fourier harmonics are compared by the $\Phi_{21}^c$, $\Phi_{31}^c$ relative phase differences (in radians, using cosine-based Fourier fits) and the $R_{21}$,  $R_{31}$ amplitude ratios, as a function of the periods ($P$). The colour coding shows the respective $R_{21}-\Phi^c_{21}$ and $R_{31}-\Phi^c_{31}$ pairs. Triangles: RRc stars, dots: RRab stars, crosses: potential anomalous Cepheid stars (see Sect.~\ref{sect:outlier}). }
   \label{fig:fourparam}
    \end{figure*}

For the \textit{K2} observations, we selected the sources near the campaign fields from the \textit{Gaia} DR2 RR~Lyrae and Cepheid classifications.
The selection of RR Lyrae candidates is described in this section, while the one related to Cepheids is discussed in Sect.~\ref{sect:cep}. Crossmatch results for both groups are listed in Appendix~\ref{app}.
We used the \texttt{K2FoV} tool to determine which stars fell on the CCD modules of \textit{Kepler} in each campaign \citep{k2fov}. Overall, 11\,361 stars were potentially observable during the selected campaigns, i.e.\ about 5\% of the RR Lyrae variables identified in \textit{Gaia} DR2. Most of these are near the Bulge or the Sgr stream (9843 stars for Campaigns 2, 7, and 11) and only 1518 fell into the halo FoVs. We crossmatched these sources with the \textit{K2} Ecliptic Plane Input Catalog (EPIC) and with the list of targets selected for observation in the mission \citep{epic}. 

The \textit{K2} observations contain various systematics caused by the excess motion of the space telescope, and multiple solutions were developed to correct for these. We generated an initial classification list by selecting the best light curves created by the various photometric pipelines. For the majority of the stars we used the official pipeline-produced Pre-Search Data Conditioned Simple Aperture Photometry (PDCSAP) products \citep{pdcsap}. Where PDCSAP was not available or was of inferior quality, we used data from the \textit{K2} Extracted Lightcurves (or K2SFF), the EPIC Variability Extraction and Removal for Exoplanet Science Targets (EVEREST), and, in a few cases, the \textit{K2} Planet candidates from OptimaL Aperture Reduction (POLAR) light curves \citep{k2sff,everest,polar}. For a few stars none of the light curve solutions provided by MAST were useful, and we applied the \texttt{PyKE} software to generate tailor-made Extended Aperture Photometry to obtain better data \citep{eap,pyke2,pyke3}. 

The distribution of the RR~Lyrae stars from the \textit{Kepler} and \textit{K2} missions crossmatched with the \textit{Gaia} DR2 candidates are plotted in Fig.~\ref{fig:rr_map}, in equatorial coordinates.

For the \textit{K2} observations, our validation procedure was based on both visual inspection and quantitative properties of the light curve shapes. 
Our criteria for the visual inspection were the following. Fundamental-mode RR Lyrae and classical Cepheid stars have very distinct, almost sawtooth-like light curves with short, sharp rising branches and long descending branches that repeat (almost) regularly, or vary smoothly if they are modulated. Characteristic bumps or humps, generated by shockwaves appearing at the surface of the star also show up at distinct pulsation phases. Hump/bump-like features can also appear in rotating variables but in these cases they drift in phase due to differential rotation. 

Double-mode stars can be harder to identify as they have less asymmetric light curves. They usually still have steeper rising branches, often with prominent humps before maximum light. Scatter in the light curve extrema may indicate beating with additional modes. In contrast, regular or smoothly varying alternations in the minima or maxima can be attributed to differences in the components and/or appearance and evolution of spots eclipsing binaries. Nevertheless, classification of nearly sinusoidal light curves can remain ambiguous. 

Overtone stars show distinct beating patterns in the light curve that, unlike beating caused by spot patterns in rotating stars, repeat very regularly. Given the small number of double-mode stars, we examined all candidates in more detail before accepting them (see below). 

For a quantitative analysis, we calculated the Fourier fits of the five strongest frequency components in the light curve using the \texttt{LCFit} code \citep{lcfit}. Then we calculated the relative Fourier parameters, comparing the $i$th and the first harmonics with amplitude ratios $R_{i1} = A_i/A_1$ and phase differences $\Phi_{i1}^c = \Phi^c_i-i\,\Phi^c_1$ (where $c$ refers to cosine-based Fourier fits), as defined by \citet{fourparam1,fourparam2}. These terms are frequently used to classify pulsating variables and to separate various subclasses. We plotted them as a function of period for $i=2,3$ in Fig.~\ref{fig:fourparam}. The stars clearly separate into the RRab and RRc loci, with very few outliers (for comparison, we refer to \citealt{soszynski2009}). We discuss some of the outliers in Sect.~\ref{sect:outlier}. The low quality of light curves in the dense stellar field of the Sgr stream prevented us to calculate Fourier parameters only in six cases. But because hints of RRab variability can be visually confirmed in them, we included these stars as positive detections.

To validate double-mode candidates, we also computed the period ratios of the two main modes of the RRd stars: these give strong constraints if the star is a normal or anomalous RRd star \citep{arrd}. These values, along with the analysis of all RRd stars observed in the \textit{K2} mission, will be published in a separate paper (Nemec, priv. comm.). 

   \begin{figure}
   \centering
   \includegraphics[width=1.0\columnwidth]{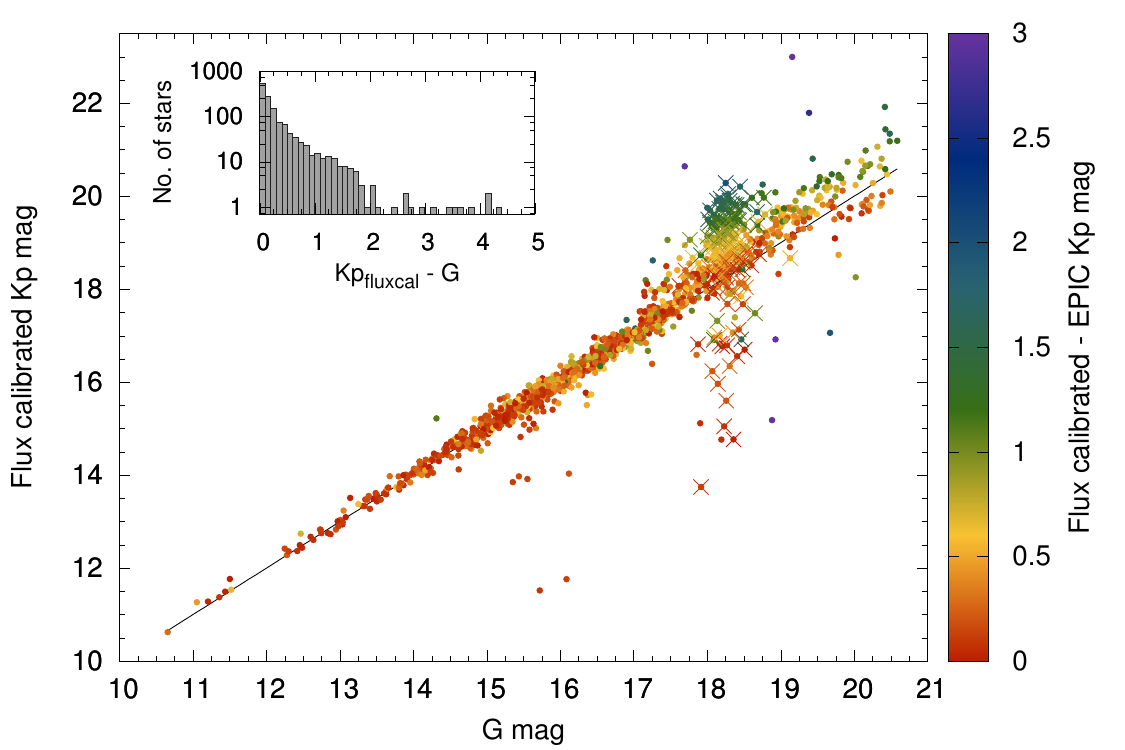}
   \caption{Comparisons of the \textit{Gaia} DR2 median $G$ magnitudes and the flux-calibrated \textit{Kp} magnitudes from the \textit{K2} mission, with the insert showing the distribution of the absolute differences. Colour coding marks the absolute differences between the flux-calibrated \textit{Kp} values and those found in EPIC: there is a clear systematic difference between the two \textit{Kp} values towards the faint end. Red outliers indicate agreement between the \textit{Kp} values but difference from the \textit{Gaia} brightnesses. Crosses mark the Sgr stream stars in Field 7. The black line marks equality. }
   \label{fig:mags}
    \end{figure}
    
\subsection{Brightness comparison}

We compared the \textit{Gaia} DR2 median \textit{G} brightnesses with the values obtained from the \textit{K2} data. The passbands of the two missions are similar, with \textit{Kepler} spanning a 420--900~nm and \textit{Gaia} a slightly wider 350--1000~nm wavelength range, both peaking around 600--700~nm \citep{keplerhandbook,evans2018}. 

We computed the flux-calibrated brightnesses (\textit{Kp}$_{\rm fc}$) of the \textit{K2} stars from the PDCSAP light curves, using the zero point of 25.3 mag, as determined by \citet{lund2015}. The two measurements generally agree well (see Fig.~\ref{fig:mags}), with 60\% of the stars below 0.1~mag difference, although it exceeds 1~mag for about 10\% of the targets.  The flux-calibrated magnitudes provide a better agreement with \textit{Gaia} than the values found in EPIC (\textit{Kp}$_{\rm EPIC}$) up to about $G\approx 18$ mag. For $G>18$, as the colour coding in Fig.~\ref{fig:mags} illustrates, the flux-calibrated values appear to be systematically fainter than the EPIC values. These issues probably come from the properties of the PDCSAP pipeline (poor background correction for dense fields and small pixel apertures for faint targets). This is especially pronounced for the population of the Sgr stream stars, marked with crosses in Fig.~\ref{fig:mags}.

For about 1\% of the stars, the \textit{Gaia} brightness differs significantly from the two nearly identical \textit{Kp} values (|\textit{Kp}$_{\rm fc}$ -- \textit{Kp}$_{\rm EPIC}|<0.1$~mag, while $|G$ -- \textit{Kp}$_{\rm fc}|>1.0$~mag). The cause of discrepancy in most of these cases is that the crossmatched EPIC actually refers to a close-by (within 1-3 \textit{Kepler} pixels), brighter star that is blended with the faint RR~Lyrae variable. In these cases, the light curve also consists of the combined flux of the two stars. Many of these \textit{G}-\textit{Kp} discrepant stars are also members of the Sgr stream.

\section{Results}
\label{sect:res}
The two quantities we are most interested in are: the purity, defined as the fraction of bona-fide variable stars of the appropriate class in the sample, and the completeness, defined as fraction of the sources (properly) identified in \textit{Gaia} DR2 compared to all (known) RR Lyrae stars within the fields of view. We investigated the distribution of stars according to the classification score, the number of \textit{Gaia} FoV transits in the $G$~band and the median $G$ brightness.

\subsection{RR Lyrae stars in the original \textit{Kepler} field}

We identified 48 \textit{Gaia} DR2 targets that were observed by \textit{Kepler} and we were able to confirm 44 of them as RR~Lyrae variables, suggesting a purity of 92\%. Beyond these 44 stars, 12 additional RR~Lyrae stars have been identified by the KASC Working Group in the Lyra-Cygnus field, not all of which are published yet, indicating a completeness of 78\% (44/56). 
One of the stars that is missing from the \textit{Gaia} DR2 RR~Lyrae sample is RR~Lyr itself (as explained in Rimoldini et al., in preparation), which was observed in the original FoV of \textit{Kepler} \citep{kolenberg2011}. (RR~Lyr is present in DR2, but with an erroneous mean $G$ brightness and parallax values \citep{brown2018}.)

After checking the data from the targeted observations, we then generated FFI-based light curves for a further 267 stars. We were able to classify 147 and 38 as (potential) RRab or RRc variables. The photometry of a further 10 targets was not successful as they were either near the edges of a CCD module or were blended with bright stars: we did not include these in our statistics. Combined with the LC targets discussed above, we can provide a lower-limit estimate of  at least 75\% for the purity of the \textit{Gaia} DR2 RR~Lyrae candidates within the original \textit{Kepler} field, although the low value can partially be attributed to the sparse FFI data we used. The completeness of the sample rises to 96\%, if we include the stars confirmed by the FFI light curves. A more detailed analysis of the FFI sample will be published elsewhere (Moln\'ar \& Hanyecz, in prep.). 

\begin{figure}
   \centering
   \includegraphics[width=1.0\columnwidth]{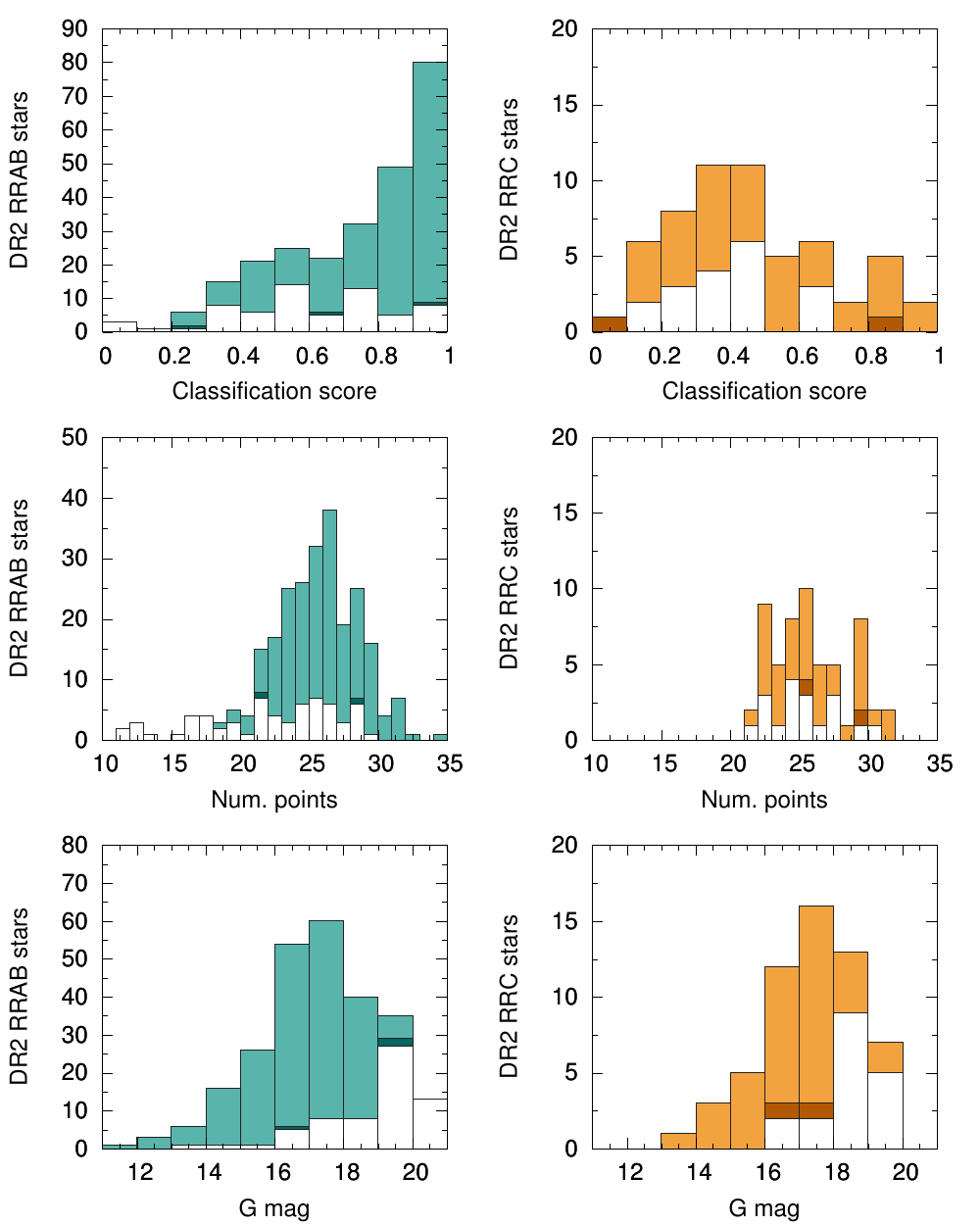}
   \caption{Histograms for the \textit{Gaia} DR2 RRAB- and RRC-classified objects in the original \textit{Kepler} field. The light hues (light-green and orange) mark stars that were confirmed as RRab or RRc variables based on the \textit{Kepler} light curves. Dark hues are RR Lyrae stars that we reclassified into the other group (dark-green: RRc variables found in the RRAB class; brown: RRab stars in the RRC class). White indicates stars that we could not confirm as RR Lyrae variables.  }
   \label{fig:keplerfield_stats}
    \end{figure}

The distribution of the stars in the original \textit{Kepler} field against the classification scores, the number of \textit{Gaia} $G$-band FoV transits, and the median $G$ brightnesses are shown in Fig.~\ref{fig:keplerfield_stats}. There is very little cross-contamination between the RRAB and RRC classes (stars that we classified into a different type). Contamination from sources that we could not confirm as RR~Lyrae variables is significant among stars that have low coverage and/or are faint ($G>18$~mag). However, this can be partially attributed to the FFI photometry pipeline that was not developed to handle very faint \textit{Kepler} targets. Purity of the FFI sample is 90\% or 85\% if we limit the targets to $G<18$ or $G<19$~mag, respectively.

\subsection{RR Lyrae stars in the \textit{K2} fields}
Overall, we were able to inspect the light curves of 1395 cross-matched \textit{K2} targets from Campaigns 0--8 and 10--13. The distribution of the \textit{Gaia} DR2 candidates observed in the K2 mission is not uniform: the 601 stars from Campaigns 2, 7, and 11 represent only 6\% of the 9843 observable targets in those fields, whereas for the rest, the halo fields the ratio is 52\% (787/1518). 

   \begin{figure}
   \centering
   \includegraphics[width=1.0\columnwidth]{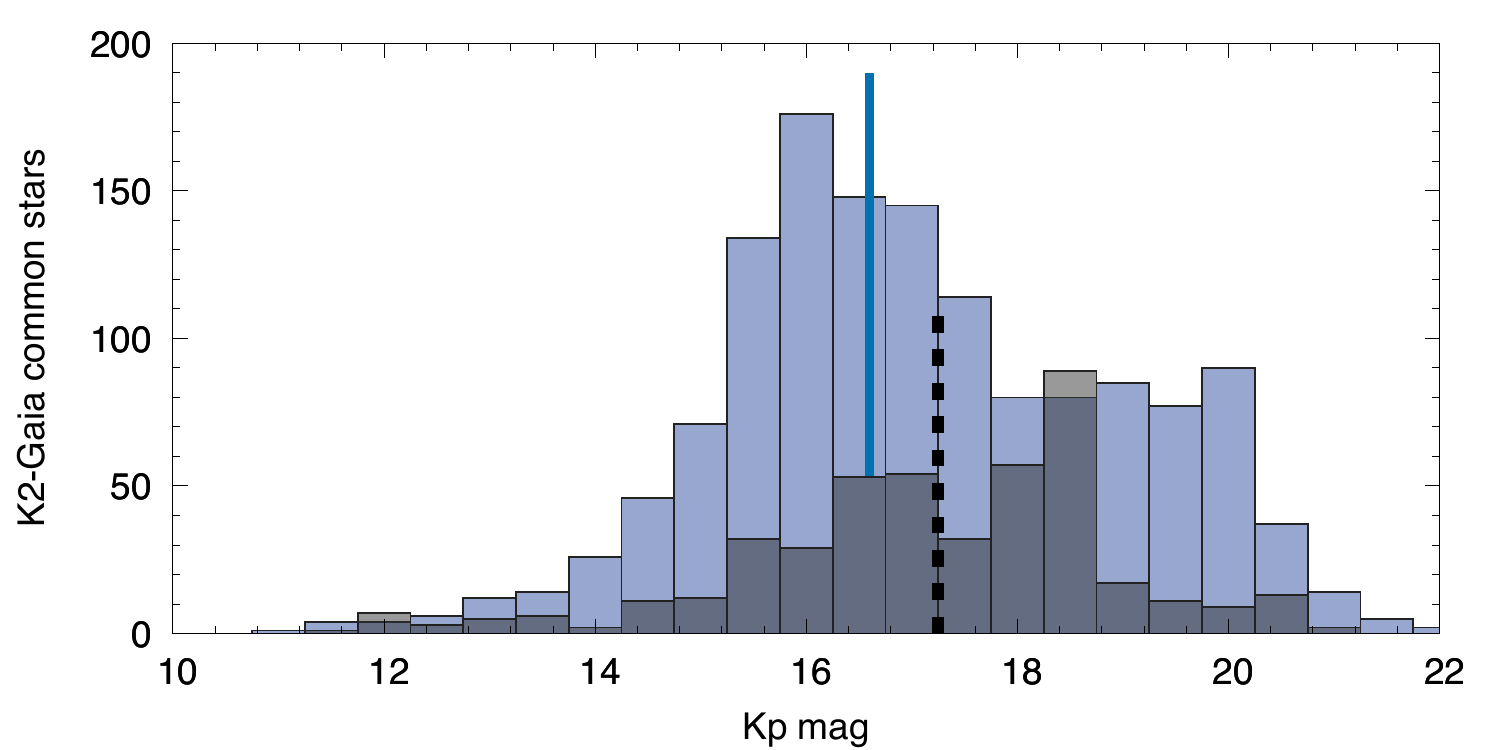}
   \caption{Brightness distributions of the K2 targets stars that are classified as RR~Lyrae variables both in \textit{Gaia} DR2 and based on their K2 light curves (blue), versus those that were missed as variable candidates in \textit{Gaia} DR2, but their K2 light curves show RR~Lyrae variation (grey). The blue solid and black dashed lines indicate the median values. Note that these histograms are overlaid, not stacked.} 
   \label{fig:notfound}
    \end{figure}

Of these 1395 stars, we confirmed the RR Lyrae-type variability of 1371 stars (1243 RRAB, 141 RRC, and 10 RRD sources). No data was available for the 5 ARRD-type stars that were within the \textit{K2} fields of view. The remaining 24 observed stars turned out to be different types of variables. These numbers lead to a \textit{Gaia} DR2 purity of 98\%, in agreement with those of the targeted observations of the original Lyra-Cygnus field, and the findings of \citet{holl2018}. However, we emphasize again that the observed stellar samples of the \textit{K2} campaigns are not necessarily representative of the true stellar populations within those fields, therefore the we treat the purity value as an upper limit. We discuss the range of possible purity values in Sect.~\ref{sect:stats} in more detail. 


Although the \textit{K2} photometric data are superior to other surveys, the observed sample is not exhaustive, so we can only provide an estimate for the completeness of the \textit{Gaia} DR2 classifications. Targets proposed for observation in the \textit{K2} campaigns were crossmatched from various surveys, and vetted based on the available photometry in the literature, and as such, are more complete than any single catalog \citep{plachy2016}. Light curves gathered by \textit{Kepler} based on these proposals were also checked visually, and they revealed very low level of contamination by other variables. Therefore we considered the number of proposed and observed stars to be a good estimate for the number of true RR~Lyrae stars in the fields of view. We then collected all stars that were proposed and observed during the mission but had no counterparts in \textit{Gaia} DR2 RR~Lyrae classifications. We ended up with 445 targets, leading to a completeness estimate of 75\% for the \textit{K2} fields. This is somewhat higher than the values computed for the OGLE fields \citep{holl2018}, but agrees with our estimate for targeted observations in the original \textit{Kepler} field. We plotted the brightness distribution of the confirmed and missed RR~Lyrae stars from the K2 mission in Fig.~\ref{fig:notfound}. The two distributions are fairly similar, but the maximum is shifted towards fainter magnitudes for the stars that are not classified as RR~Lyrae in the \textit{Gaia} DR2 classification table, by about 0.6 mag (the medians of the two groups are 16.6 and 17.2 mag).

{We also compared the brightness distributions of all \textit{Gaia} DR2 candidates falling into the K2 fields to all confirmed RR Lyrae stars therein (including the missed 445 stars) in Fig.~\ref{fig:selfunc} to see how different the selection function of the two missions are. Based on their capabilities alone, \textit{Kepler} is only limited by source confusion but it is able to observe stars below the faint limit of \textit{Gaia} (see, e.g., the RR Lyrae stars in Leo IV, \citealt{molnar2015}). However, the sample observed by \textit{Kepler} was limited by the input catalogs used for target selection. The large difference in the upper panel of Fig.~\ref{fig:selfunc} comes from the large number of bulge stars that were not observed in the K2 mission. The comparison of the halo fields only (i.e., excluding Campaigns 2, 7, and 11) shows a much better agreement in the lower panel. For stars brighter than 16.5 mag (here we used either $G$ or \textit{Kp} magnitudes for stars, given the good agreement between the two) the K2 observations and the input catalogs we used provide a sample more complete than that of \citet{holl2018}. Interestingly, the \textit{Gaia} DR2 sample shows another excess below 18.5 mag. Since most of these stars have no K2 light curves, we cannot decide if these stars are contaminants or \textit{bona fide} RR Lyrae stars that were not detected by other surveys before. Unfortunately, this brightness range will not be accessible to the TESS space telescope either. 

   \begin{figure}
   \centering
   \includegraphics[width=1.0\columnwidth]{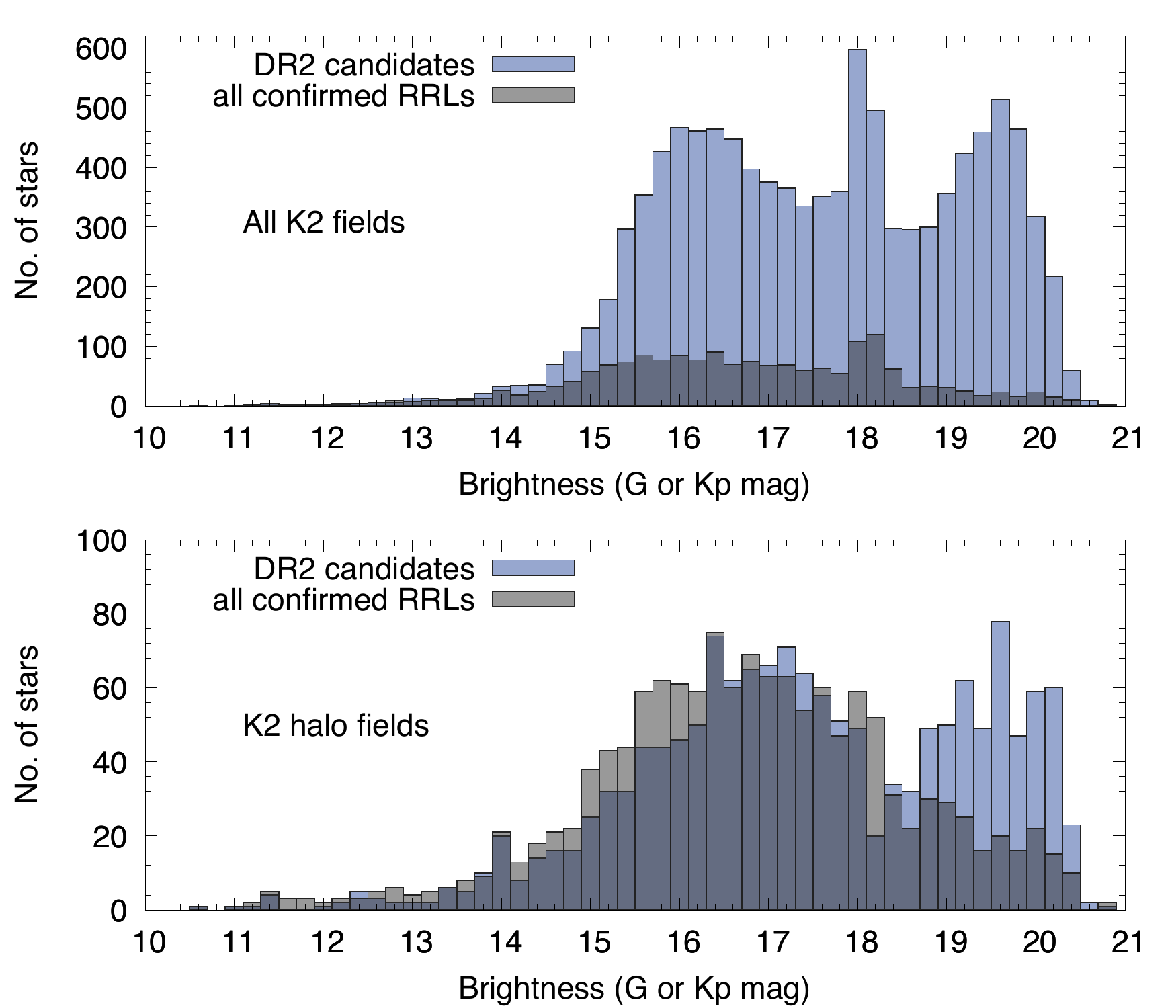}
   \caption{Brightness distributions of the \textit{Gaia} DR2 candidates (blue) by \citet{holl2018} in the K2 fields, versus all known, confirmed RR Lyrae stars within in the same fields, including ones not in the DR2 variability catalogg. The upper panel shows all fields, the spike at 18 mag is the Sgr stream. The lower panel shows the halo fields only.   }
   \label{fig:selfunc}
    \end{figure}

\subsubsection{RR Lyrae subclass statistics}
Based on their K2 light curves, we identified 1371 objects as RR Lyrae stars, 1211 (88\%) RRab, 142 (10\%) RRc, 17+1 RRd and anomalous RRd stars (1\%). Our classifications do not always agree with those of \textit{Gaia} DR2. We found thr \textit{Gaia} DR2 RRAB class to be nearly pure, with only 1\% (14/1243) of contamination from the other classes (stars that turned out to be RRc- or RRd-type pulsators instead). The contamination in the RRC and RRD classes were 8\% (12/142) and 50\% (5/10), respectively. While contamination rises significantly for these classes, they are much less numerous, therefore the overall  rate is only about 2\% for the RR Lyrae stars in general.

Moreover, out of the 24 stars we could not confirm as RR Lyrae variables, \textit{K2} light curves of 5 RRAB candidates revealed Cepheid variations (1~anomalous and 4 Type II Cepheids). This indicates a low level of cross-contamination between the RR Lyrae and Cepheid classes. We reclassify two of these, V1637~Oph (EPIC~234649037, $P_\mathrm{K2}=1.327$~d), previously classified as an RR~Lyrae, and FZ~Oph (EPIC~251248334, $P_\mathrm{K2}=1.500$~d), an under-observed variable, as short-period Type II Cepheids, also known as BL~Her-type stars, based on their \textit{K2} light curves. The results are summarized in the confusion matrix in Fig.~\ref{fig:confmatrix}.

\begin{figure}
   \centering
   \includegraphics[width=1.0\columnwidth]{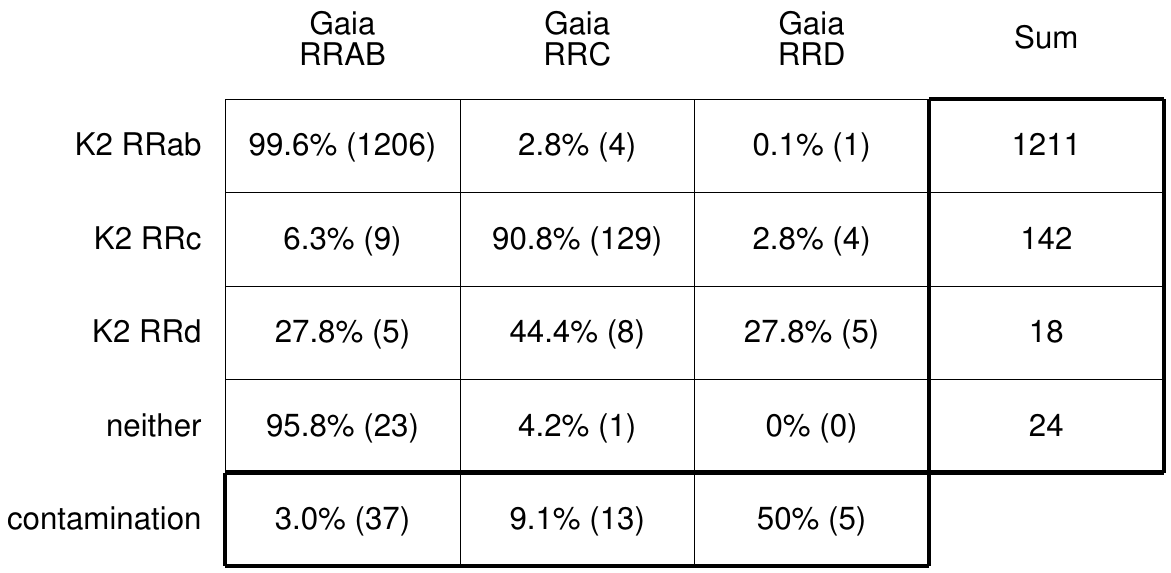}
   \caption{Confusion matrix of the \textit{Gaia} and K2 results. The \textit{Gaia} classifications are compared to our findings. Contamination of each \textit{Gaia} class is presented in the bottom row. }
   \label{fig:confmatrix}
    \end{figure}

Fig.~\ref{fig:scorebin} shows the distribution of the classification scores for the various RR~Lyrae subclasses. The left panels group the stars according to their \textit{Gaia} DR2 classification types. We indicated the portion of stars we reclassified \textit{from} these classes with darker hues. The right panels show the variability types based on the \textit{K2} light curves, with the same colours denoting the number of stars that we reassigned \textit{to} a different RR~Lyrae subclass. The difference between the classification score distributions of RRab and RRc/RRd stars is striking: half of the RRab stars have scores above 0.8, while the distribution of the RRc and RRd stars is essentially flat, and for RRc stars it falls off near~1.0. The flat distribution suggests that the RRC and RRD classes were harder to identify, likely due to the competition with the dominant RRAB class. Figure~\ref{fig:scorebin} also confirms that the RRab sample is nearly pure (almost all stars are from the RRAB class), while the RRd sample includes \textit{Gaia} classifications of all three subclasses. For the RRc stars, the sample appears to be nearly pure above classification score 0.6, and about 15\% of RRc stars with scores $<0.6$ ended up in the RRAB or RRD classes of the \textit{Gaia} DR2 classification. 

   \begin{figure}
   \centering
   \includegraphics[width=1.0\columnwidth]{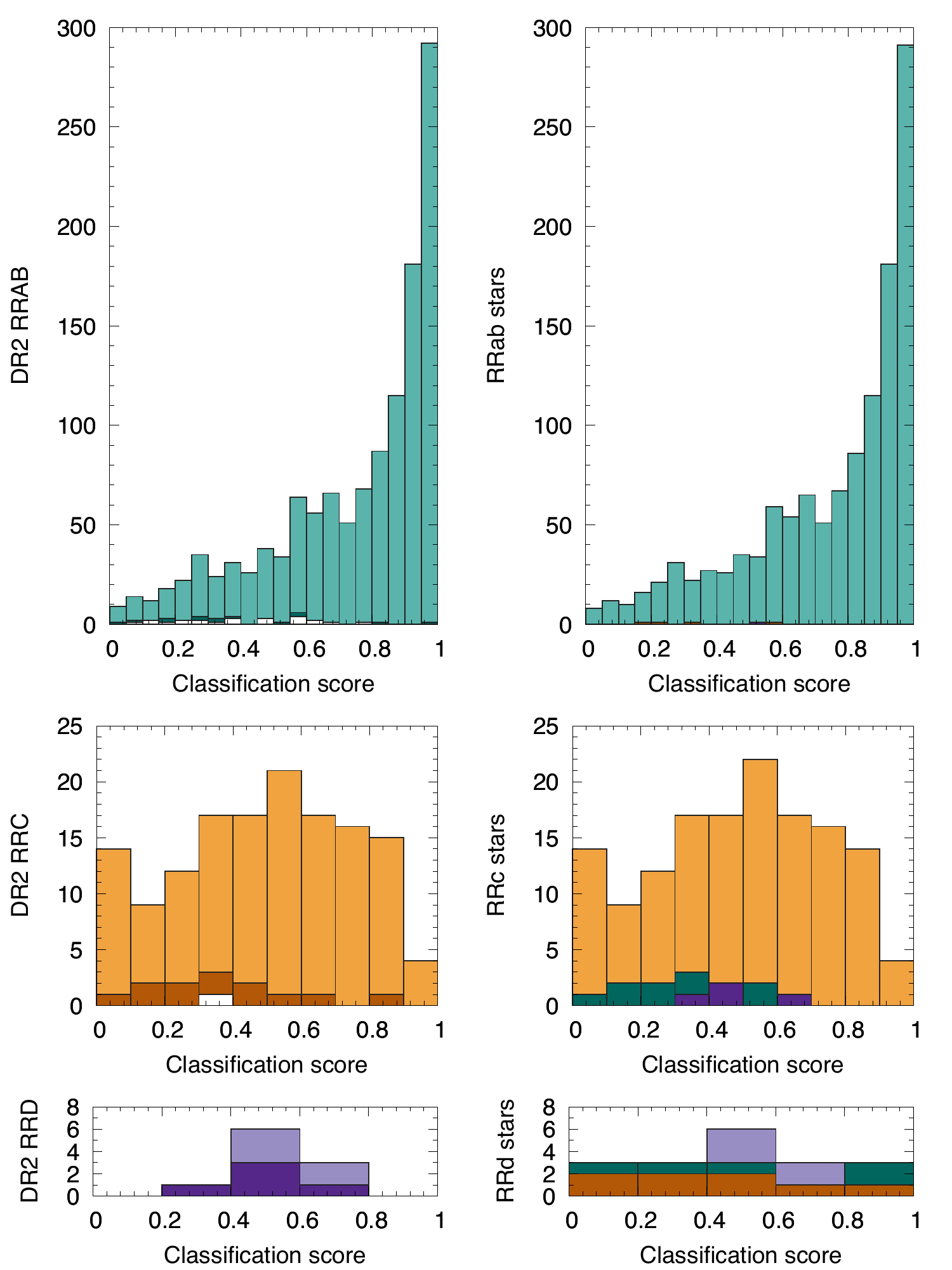}
   \caption{Distribution of the various classification scores of the RR Lyrae subtypes. Left: \textit{Gaia} DR2 classification type; right: as classified based on the \textit{K2} data. The distribution is skewed towards score values of 1.0 for RRab stars but is flat for RRc and RRd stars. Light green, orange and light lilac colours mark bona-fide RRab, RRc, RRd stars, respectively. Dark green, brown and dark lilac colours denote stars that we classified into a different subclass (which are distributed in different panels on the right-hand side). White bars refers to the counts of stars for which the RR Lyrae variability was not confirmed or was rejected.  }
   \label{fig:scorebin}
    \end{figure}

   \begin{figure}
   \centering
   \includegraphics[width=1.0\columnwidth]{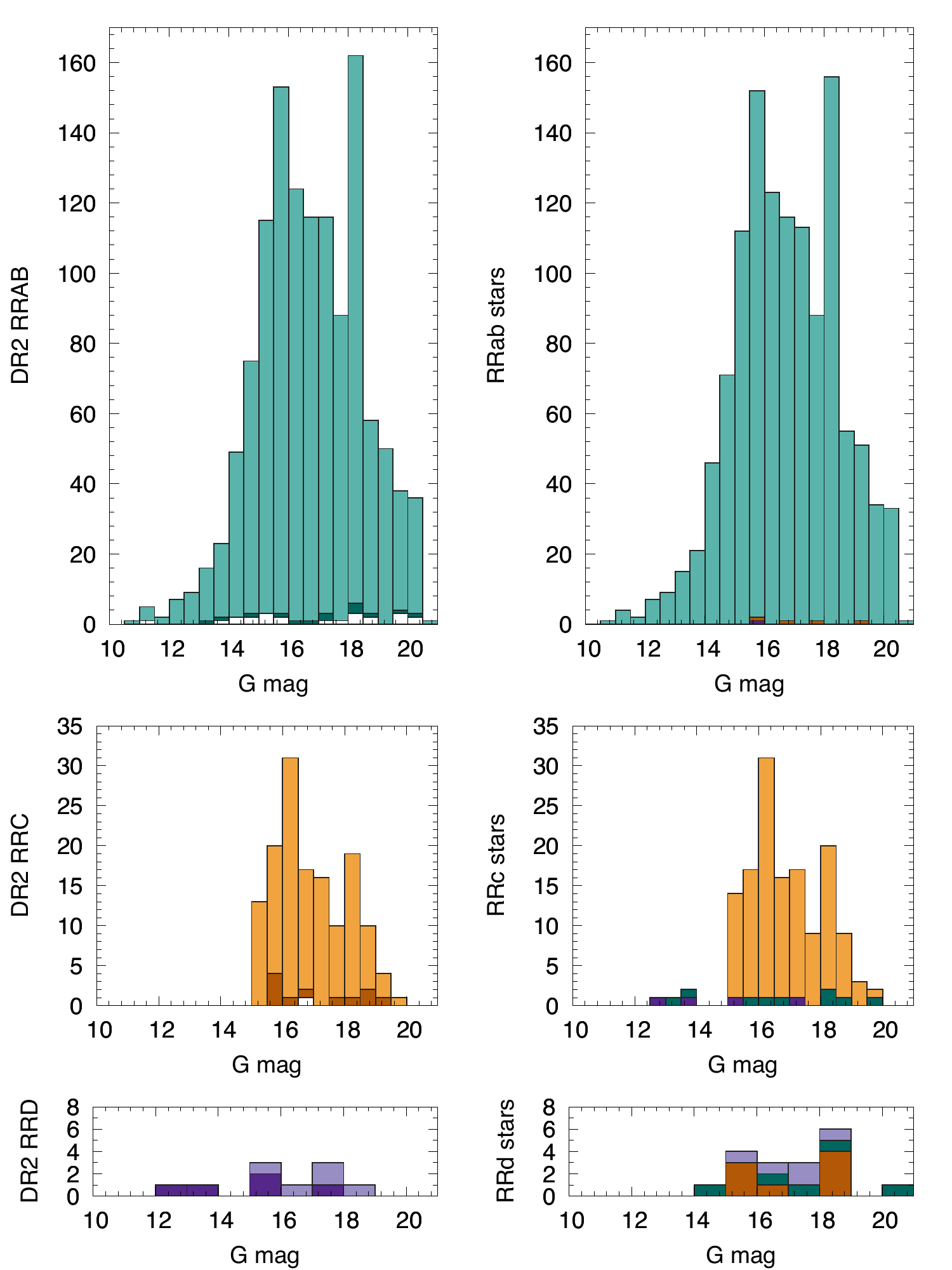}
   \caption{Distribution of the median $G$ magnitudes per RR~Lyrae subtype. The colour coding is the same as in Fig.~\ref{fig:scorebin}.}
   \label{fig:magbin}
    \end{figure}

   \begin{figure}
   \centering
   \includegraphics[width=1.0\columnwidth]{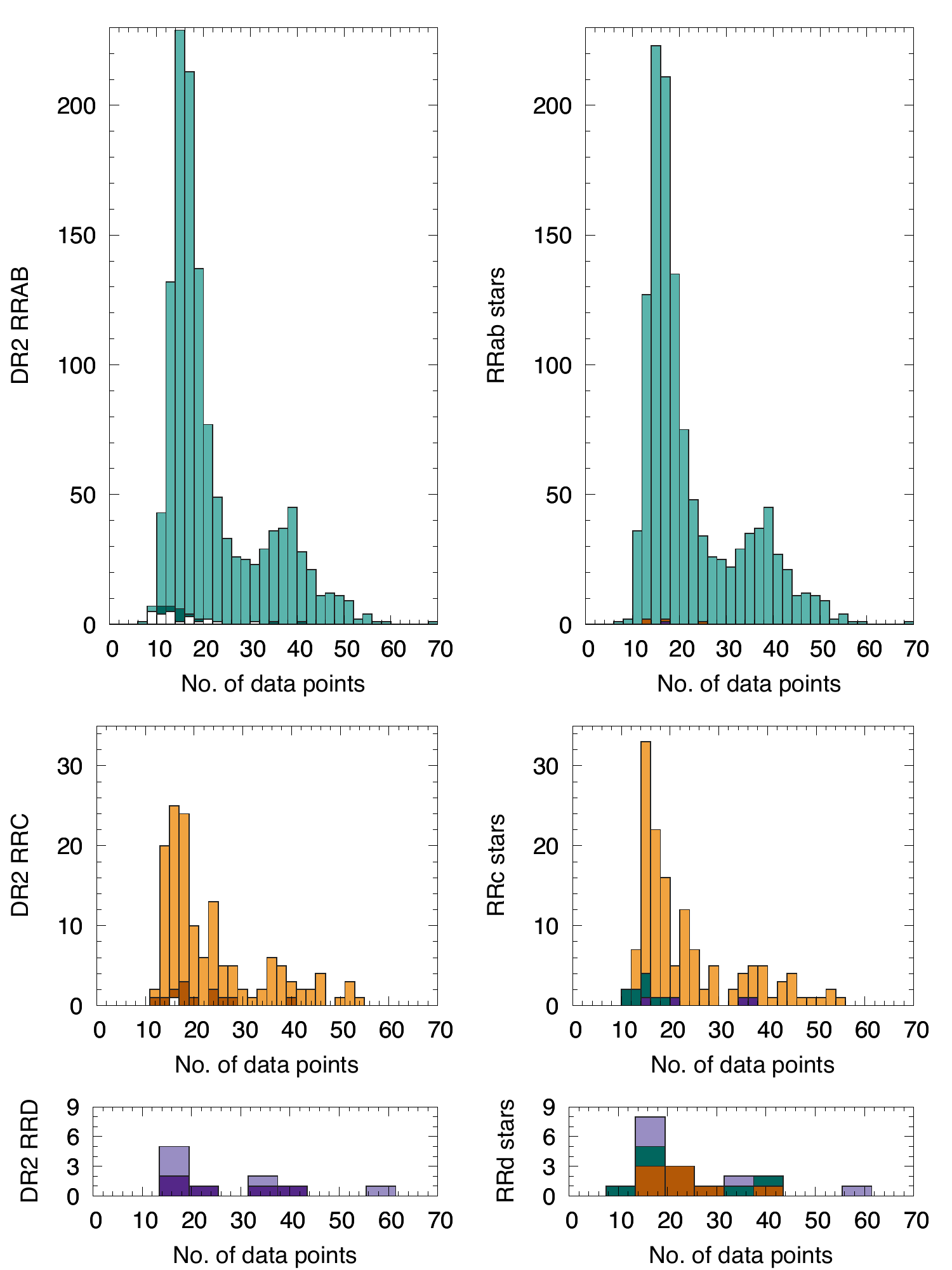}
   \caption{Distribution of the number of field-of-view transits in the $G$ band per RR~Lyrae subtype. The colour coding is the same as in Fig.~\ref{fig:scorebin}.  }
   \label{fig:numbin}
    \end{figure}

The distributions of the median brightnesses in the $G$ band of the three RR~Lyrae subclasses are shown in Fig.~\ref{fig:magbin}. Contrary to the  distribution from the original \textit{Kepler} field, we do not see an increase of unconfirmed variables towards the faint end, as they appear rather evenly spread. Interestingly, RRc stars brighter than $G\sim 15$~mag seem to be missing from our particular selection of \textit{Gaia} DR2 classifications (centre-left panel), although we reclassified a few of them from the other subclasses  (centre-right panel). However, the number of RRab stars also decreases rapidly from 14-15 mag towards the bright end. Therefore the apparent lack of a few RRc stars most certainly stems from small number statistics here. 

Finally, Fig.~\ref{fig:numbin} shows the distribution of stars according to the number of $G$-band observations, per RR~Lyrae subclass. Here, the unconfirmed variables clearly group at the low end, especially for the RRAB class. For stars with 20~FoV transits or less, the contamination is about 3\%, and for the few stars below 12~FoV transits, it rises to 8\%. These numbers indicate that the \textit{Gaia} DR2 sample of RR~Lyrae classifications is relatively pure even for stars with few observations. 

These estimates illustrate that the classifications of \textit{Gaia} DR2 are already well-suited to identify single-mode pulsators, even at low numbers of FoV transits, but multiperiodic objects like RRd stars will need more extended observations. 

\subsubsection{Ambiguous identifications}
\label{sect:outlier}
As mentioned in Sect.~\ref{sect:id}, the Fourier parameters in Fig.~\ref{fig:fourparam} mostly separate the stars into two groups corresponding to RRab- and RRc-type stars, but we found a few stars that we could not classify unambiguously. We flagged four stars as potential anomalous Cepheids: three fundamental-mode and one first-overtone candidates. But we must emphasize that the Fourier parameters of anomalous Cepheids overlap with those of RRab stars for certain period ranges, and therefore we could not rule out that these four objects are RRab stars based on their light curve shapes alone. These four stars are identified in \textit{K2} as EPIC~206010651, 206175324, 212459957, and 234523936 (the last one is the overtone candidate), corresponding to \textit{Gaia} DR2 \texttt{source\_id} 2600030303142307968, 2614960399737036672, 3606980678405498368, and 4134356134978875904, respectively.

Some outliers of the distribution of $\Phi^c_{21}$ in Fig.~\ref{fig:fourparam} can be attributed to a peculiar group of modulated RRab stars that exhibit a strong Blazhko effect, with a full rotation in the $\Phi^c_{i1}$ parameters during the minimum-amplitude cycle \citep{guggenberger2012,bodi2018}. This phenomenon could shift the average value of the parameters when the duration of the light curve is comparable to the modulation period. This is the case, for example, for EPIC~245954410, with $\log P/\mathrm{d}=-0.29$ and $\Phi^c_{21}=6.16$~rad, and for EPIC~212545143, an extremely modulated RRab star with $\log P/\mathrm{d}=-0.27$ and $\Phi^c_{21}=0.99$~rad. We note that two other stars at low $\Phi^c_{21}$ but with shorter periods are not flagged as outliers. Although most RRc stars appear in a tight group, others spread out from 0 to $2\pi$, and these two stars represent that subgroup of RRc stars.

\subsection{RR Lyrae completeness and purity ranges}
\label{sect:stats}
The observations of the two missions of \textit{Kepler} provide different samples that are not straightforward to combine. Here we go through the various samples and statistics obtained. One can focus on the original \textit{Kepler} field where the combination of the targeted, continuous light curves and the FFI photometry provides a sample that is not biased by prior selection of targets, but hindered by the sparse nature of the FFI light curves. Alternatively, one can gather all targeted observations from the two missions resulting in a larger sample, based on the same data acquisition method. Finally, we can calculate purity and completeness values for all three samples (\textit{Kepler}, \textit{Kepler} FFI, and \textit{K2}) combined. Through these combinations we provide a range of estimates, which are summarized in Table~\ref{tab:results}. 

Overall, we were able to identify 0.9\% of the RR~Lyrae variables classified by Rimoldini et al.\ (in preparation) over the whole sky in the observations of the two missions of \textit{Kepler}, outside the Bulge. Inside the \textit{Kepler} and K2 fields of view, 15\% of the DR2 targets had light curves collected. (1706 stars out of 11\,702 potential, observable targets from the two missions.) But, as stated above, the actual coverage is much higher for the original \textit{Kepler} FoV (near-complete) and the halo fields in the \textit{K2} mission (52\%).


For the \textit{Kepler} and/or \textit{K2} targeted observations, we derived a purity rate between 92--98\% (with lower limits at 51\% and 75\%), and a completeness in the range of $70-78\%$. The completeness is estimated by simply comparing the number of crossmatched and confirmed \textit{Gaia} sources to those also in the fields but not classified as RR~Lyrae variables in \textit{Gaia} DR2. The lowest completeness value of 70\% here is reached if we only use the targeted observations but count the FFI stars among the missed ones (even though they were not known before DR2). The DR2 classifications have a completeness of 75\% within the \textit{K2} fields, while the combination of all data sets leads to an overall completeness rate of 78\%. 

We obtained a very high completeness ratio of 96\% for the original \textit{Kepler} field, if the FFI stars are included. The lack of further sources here can be attributed to the fact that the field is at low Galactic latitudes and thus was largely avoided by deep surveys that could have identified faint RR~Lyraes there. Therefore, the high completeness in this field reflects our limited knowledge that has now been expanded by the \textit{Gaia} DR2 identifications, so we consider this value as an upper limit.  

Even more diverse limits can be derived for the purity. The estimate of 92\% originates from the targeted \textit{Kepler} observations, but that sample alone is much smaller than the rest. Both the \textit{K2} and combined \textit{Kepler}+\textit{K2} observations show 98\% purity. This is in broad agreement with the findings of \citet{holl2018}. We concluded that the single-mode group RRAB is nearly pure whereas the RRC group suffers from some contamination, but most of the confusing sources are RRd stars, and not eclipsing binaries. \textit{Gaia} DR2 all-sky classifications are not well-suited yet to identify double-mode (RRd) stars, but as those are intrinsically rare, they have little effect on the overall population statistics. 

The inclusion of the FFI stars into the statistics of the original field, however, warns us that the \textit{Kepler} and \textit{K2} samples might include some bias against contaminating sources in \textit{Gaia} DR2 that have not been proposed for observation in the \textit{Kepler} and \textit{K2} missions. The combined statistics from the original field gives us a lower limit of 75\% for purity, although this value is likely affected by faint RR~Lyrae sources whose FFI photometry was of insufficient quality. 

Of course, the \textit{Kepler}+FFI sample is disjunct from the \textit{K2} FoVs, but it is the only one without prior selection bias. The excess of faint unobserved stars in the selection function (Fig~\ref{fig:selfunc}) suggests that the DR2 RR~Lyrae sample might potentially be less pure than our statistics suggests. In absence of those light curves, only a simple lower limit can be estimated for \textit{K2}, considering that in halo fields 52\% of the \textit{Gaia} sources were observed with 98\% purity. Even if all the remaining 48\% were false positives, this sets an absolute lower limit of 51\% for the purity in the Galactic halo. The combination of halo and original-field stars raises that limit slightly to 55\%.

\begin{table}
\caption{Completeness and purity values for the RR Lyrae and Cepheid stars in \textit{Gaia} DR2, based on the \textit{Kepler} and \textit{K2} light curves. Absolute lower limits of purity rates (from the worst-case scenario in the halo \textit{K2} fields) are indicated in brackets.} 
\label{tab:results}
\centering 
\begin{tabular}{l c c c c} 
\hline
\hline 
 Type & Data & \textit{Gaia} matches & Compl. & Purity \\
 \hline
 RRL & \textit{Kepler} & ~~~~48 & 78\% & 92\% \\
 RRL & \textit{Kepler} + FFI & ~~311 & <96\% & >75\% \\
 RRL & \textit{K2}     & 1395 & 75\% & (51--)98\% \\
 RRL & \textit{Kepler} + \textit{K2} & 1443 & 70--76\% & (51--)98\% \\
 RRL & All & 1706 & 78\% & (55--)94\% \\
 \hline
 CEP & All & ~~~~41 & & $\sim 66\%$ \\
 \hline
\end{tabular}
\end{table}

\subsection{Cepheids in the \textit{Kepler} and \textit{K2} data}
\label{sect:cep}
Cepheids are much less numerous than RR~Lyrae stars, and many of them populate the Galactic Bulge, the Magellanic Clouds, and the disk of the Milky Way. The \textit{Kepler} and \textit{K2} missions largely avoided these areas to prevent source confusion and blending, therefore the overlap between these missions and the Cepheids in \textit{Gaia} DR2 is much more limited than for the RR~Lyrae stars. Because of the low number of targets, we decided to also include the Bulge stars in our statistics. Since Cepheids are intrinsically luminous, their variations can be recognized even when blended with other stars in the \textit{Kepler} observations.

In the Lyra-Cygnus field of the original \textit{Kepler} mission, \textit{Gaia} DR2 classifications correctly include V1154~Cyg as CEP and HP~Lyr as T2CEP \citep[which is a potential RV~Tau star,][]{hplyr}, while it missed DF~Cyg and misclassified V677~Lyr as T2CEP 
\citep[known to be a longer-period semiregular variable,][]{v677lyr}. 

In the \textit{K2} fields, 73 stars classified as Cepheids in \textit{Gaia} DR2 were observable. Of these, we found data for 38 from the three subclasses (3 ACEP, 13 CEP, and 22 T2CEP stars), of which we were able to confirm 20 as a members of the Cepheid family, and 5 more as likely Cepheid stars. Given the small sample size, we did not separate the stars further into subclasses. With the inclusion of the stars from the RRAB class that we classified as Cepheids, the number of Cepheid variables common between \textit{Kepler/K2} and \textit{Gaia} is 32 stars (2 from \textit{Kepler} and 25+5 from \textit{K2}). 

The stars we were not able to confirm as Cepheids include 4~eclipsing binaries and/or rotational variables, 5 long-period variables, and 4 stars with unidentified variations. We classified 1~star as an RRab variable, further confirming a low level of cross-identification between Cepheids and RR Lyrae stars. This star was included in a \textit{K2} proposal as an RR~Lyrae target, and thus it has been accounted for in our completeness estimate. 


We were able to inspect the light curves of 54\% (41/76) of the stars that fell into the fields of view of the two missions. The stars identified in the two \textit{Kepler} missions account for only 0.9\% (observable) and 0.5\% (observed) of the 8550 Cepheid variables identified by Rimoldini et al.\ (in preparation). However, most of the DR2 detections concentrate in the Magellanic Clouds, therefore our results are more relevant for the Cepheid population in the Milky Way. The findings indicate a purity in the order of 66\% for this size-limited sample. Considering the very low number of sources available to us, and the limited temporal coverage of the \textit{K2} data, this result is more or less in agreement with the 15\% contamination in the sky-uniform test results of \citet{holl2018}. 

   \begin{figure}
   \centering
   \includegraphics[width=1.0\columnwidth]{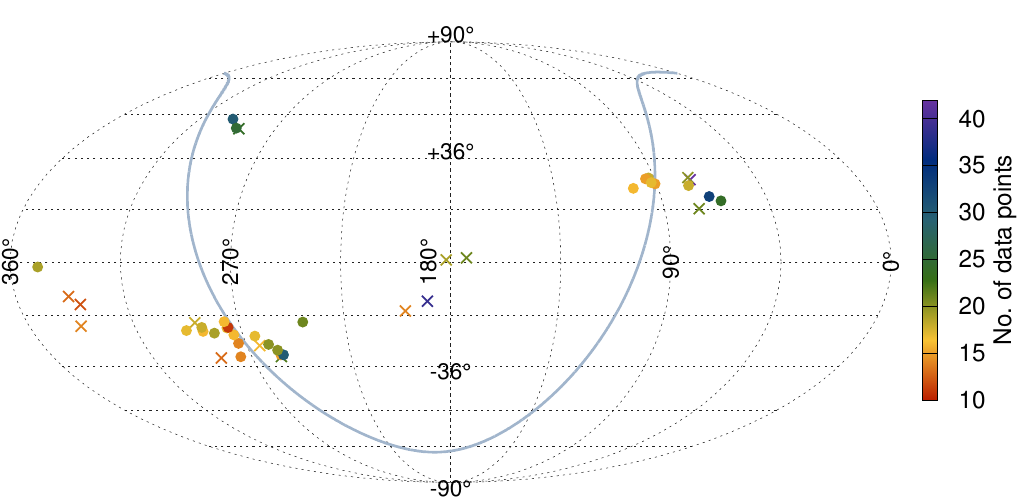}
   \caption{The distribution in the sky of various Cepheid-type stars from the \textit{Gaia} DR2 classifications crossmatched with \textit{Kepler} and \textit{K2} measurements. Crosses mark stars the we rejected or were not able to confirm: half of these are at high Galactic latitudes. The notation is the same as in Fig.~\ref{fig:rr_map}. }
   \label{fig:cep_map}
    \end{figure}

\section{Conclusions}
\label{sect:concl}
RR~Lyrae and Cepheid stars are used for many purposes, such as mapping the substructures and stellar populations of the Milky Way and the Magellanic Clouds. The \textit{Gaia} Data Release 2 features a large collection of RR~Lyrae and Cepheid  candidates (among others), over the entire sky that can be exploited for such studies. However, many of the identifications are based on a low number of observations. The \textit{Kepler} space telescope observed a selection of these targets in great detail during the \textit{Kepler} and \textit{K2} missions. The fact that \textit{Kepler} is able to provide continuous light curves for the entire brightness range of the \textit{Gaia} DR2 targets provides a great opportunity to validate these classifications. We investigated the targets in common between the missions, consisting of 0.9\% of both the RR~Lyrae and Cepheid candidates from the all-sky classification of DR2 (Rimoldini et al., in preparation). Within the areas of the \textit{Kepler} and \textit{K2} fields of view (between Campaigns 0-8 and 11-13), 15\% and 54\% of the DR2 RR~Lyrae and Cepheid candidates, respectively, had LC data recorded by \textit{Kepler}. For the original field of view we also extracted sparse photometry from the Full Field Images, to obtain a sample more complete than the targeted observations alone. 

We found that the photometry in \textit{Gaia} DR2 is already suitable to properly identify single-mode pulsators. The RRAB class was found to be nearly pure, with very few contaminants. The RRC class and the various Cepheid classes are somewhat more contaminated. The few RRD (double-mode) stars we found included single-mode pulsators as well, and bona-fide RRd stars have \textit{Gaia} DR2 classifications from all RR~Lyrae subclasses, indicating that reliable identification of multimode pulsators will require more observations. Overall, we found the purity to be in the 92--98\% range, based on the targeted observations of the \textit{Kepler} space telescope, with lower limits of 75\% (FFI stars) and 51\% (worst-case scenario for the halo \textit{K2} fields). For the classification of Cepheids in \textit{Gaia} DR2, we provide a purity estimate in the order of 66\%.

Based on the visual examination of the contaminating sources (stars that were not found to be RR~Lyrae or Cepheid stars), we concluded that contaminants were more likely to be pulsators or rotational variables than eclipsing binaries. 

We estimated the completeness of the \textit{Gaia} DR2 RR~Lyrae classifications in the \textit{Kepler} and \textit{K2} fields to be around 70-78\% for the targeted observations with an upper limit of 96\% for the original field, if we included the FFI stars as well. 


All of the estimates presented here are summarized in Table~\ref{tab:results} and they are in agreement with the limited validation tests presented in \citet{holl2018}, 
indicating that the observations of the \textit{Kepler} space telescope can indeed be used to validate surveys that collect sparse photometry.  

\begin{acknowledgements}
This research received funding from the Hungarian Academy of Sciences through the Lend\"ulet grants LP2014-17 and LP2018-7, and the J\'anos Bolyai Research Scholarship (L.M.\ and E.P.); 
the Hungarian National Research, Development, and Innovation Office through the NKFIH grants K-115709, PD-116175, and PD-121203; and the \'UNKP-17-3 program of the Ministry of Human Capacities of Hungary (\'A.L.J.). L.M. has been supported by the Premium Postdoctoral Research Program of the Hungarian Academy of Sciences. This work has made use of data from the European Space Agency (ESA) mission \textit{Gaia} (\url{https://www.cosmos.esa.int/gaia}), processed by the \textit{Gaia} Data Processing and Analysis Consortium (DPAC, \url{https://www.cosmos.esa.int/web/gaia/dpac/consortium}). Funding for the DPAC has been provided by national institutions, some of which participate in the Gaia Multilateral Agreement, which include, for Switzerland, the Swiss State Secretariat for Education, Research and Innovation through the ESA Prodex program, the ``Mesures d'accompagnement", the ``Activit\'{e}s Nationales Compl\'{e}mentaires", and the Swiss National Science Foundation.
This paper includes data collected by the \textit{Kepler} mission. Funding for the \textit{Kepler} and \textit{K2} missions are provided by the NASA Science Mission Directorate. The \textit{Kepler/K2} data presented in this paper were obtained from the Mikulski Archive for Space Telescopes (MAST). STScI is operated by the Association of Universities for Research in Astronomy, Inc., under NASA contract NAS5-26555. Support for MAST for non-HST data is provided by the NASA Office of Space Science via grant NNX09AF08G and by other grants and contracts. We Dr.~Carolyn Doherty for proofreading the manuscript, and Dr.~Benjamin Montet for the fruitful discussions about the \texttt{f3} code.
\end{acknowledgements}


\bibliographystyle{aa} 
\bibliography{gaiakepler}

\appendix

\section{\textit{Gaia} DR2 and KIC/EPIC crossmatch }
\label{app}
In the following tables, we provide the crossmatch of the sources in common between \textit{Gaia} DR2 classifications and the \textit{Kepler} and \textit{K2} observations. Tables \ref{tab:krrabxmatch}, \ref{tab:krrcxmatch}, and \ref{tab:kuncertxmatch} list stars from the original \textit{Kepler} FoV that we classified as RRab, RRc, or likely RRab/RRc variables. Tables \ref{tab:rrabxmatch}, \ref{tab:rrcxmatch}, and \ref{tab:rrdxmatch} include the RRab, RRc, and RRd stars identified in the \textit{K2} observations, respectively. Longer tables are available online in full length. Table \ref{tab:cepxmatch} lists the crossmathed Cepheid-type stars. Finally, stars that we rejected as RR Lyrae or Cepheid stars based on their \textit{K2} data are presented in Tables \ref{tab:nonrrmatch} and \ref{tab:noncepmatch}.

\begin{table*}
\caption{Crossmatch of RRab-type stars in the original \textit{Kepler} observations. The \textit{Kp} brightness here refers to the values found in KIC, data type is either target (LC data available) or FFI (no LC data, just FFI photometry). The following information was extracted from the \textit{Gaia} DR2 archive: RA and Dec coordinates (\texttt{ra} and \texttt{dec} fields of the \texttt{gaiadr2.gaia\_source} table), the median \textit{G}-band magnitudes (rounded values from those in the \texttt{median\_mag\_g\_fov} field of the \texttt{gaiadr2.vari\_time\_series\_statistics} table), and the DR2 class (content of the \texttt{best\_class\_name} field of the \texttt{gaiadr2.vari\_classifier\_result} table). The full table below is available electronically. } 
\label{tab:krrabxmatch}
\centering 
\begin{tabular}{r c c c c c c c} 
\hline
\hline 
\noalign{\vskip 3pt}
\textit{Gaia} DR2 source\_id & DR2 RA & DR2 Dec & \textit{G}  & DR2 class & KIC & \textit{Kp}  & data type\\
 & (deg) & (deg) & (mag) &  &  & (mag) & \\
 \hline 
\noalign{\vskip 3pt}
2051319231466329600 & 290.0443778 & 38.2875548 & 15.852 & RRAB & 3111002 & 15.989 & FFI\\
2051756764073824512 & 291.5422064 & 37.2395927 & 18.121 & RRAB & 1721534 & 16.913 & FFI\\
2051849058633285632 & 292.8211913 & 37.6989397 & 16.471 & RRAB & 2309247 & 16.387 & FFI\\
2051927669415635072 & 292.6072021 & 38.2211264 & 16.179 & RRAB & 3121676 & 15.959 & FFI\\
2051930903531763072 & 292.5824442 & 38.2520216 & 16.165 & RRAB & 3121566 & 16.400 & FFI\\
2052112730962078080 & 294.4600661 & 38.2577522 & 17.047 & RRAB & 3129996 & 17.002 & FFI\\
2052124653792193280 & 294.5614553 & 38.4515596 & 17.952 & RRAB & 3354775 & 17.637 & FFI\\
2052162282013412992 & 295.0290071 & 38.9723355 & 15.410 & RRAB & 3864443 & 15.593 & target\\
2052226633499326336 & 294.5504877 & 39.1471277 & 14.583 & RRAB & 4069023 & 14.650 & FFI\\
2052259898026804480 & 293.0292280 & 38.4096197 & 17.767 & RRAB & 3348493 & 17.827 & FFI\\
\multicolumn{8}{l}{$\ldots$}\\
\hline 
\end{tabular}
\end{table*}

\begin{table*}
\caption{Crossmatch of RRc-type stars in the original \textit{Kepler} observations.  } 
\label{tab:krrcxmatch}
\centering 
\begin{tabular}{r c c c c c c c} 
\hline
\hline 
\noalign{\vskip 3pt}
\textit{Gaia} DR2 source\_id & DR2 RA & DR2 Dec & \textit{G}  & DR2 class & KIC & \textit{Kp}  & data type\\
 & (deg) & (deg) & (mag) &  &  & (mag) & \\
 \hline 
\noalign{\vskip 3pt}
2053453246098423424 & 291.2093828 & 40.2449799 & 14.652 & RRC & 5097329 & 14.896 & FFI\\
2073611761018174208 & 298.7963502 & 40.6390445 & 17.183 & RRC & 5476906 & 17.125 & FFI\\
2076328310640589312 & 296.0172668 & 39.9738297 & 17.391 & RRC & 4852066 & 17.393 & FFI\\
2076789487048365952 & 296.9555846 & 41.1889926 & 17.642 & RRC & 5895400 & 17.796 & FFI\\
2079302622726841728 & 298.4694906 & 45.0157871 & 16.093 & RRC & 8839123 & 16.229 & FFI\\
2080332143568818304 & 296.6246981 & 46.5426500 & 17.569 & RRC & 9783052 & 17.730 & FFI\\
2080566717498680448 & 296.7116973 & 47.2295668 & 16.997 & RRC & 10221234 & 17.050 & FFI\\
2086448009499589376 & 298.3179484 & 47.7771822 & 15.909 & RRC & 10553801 & 15.700 & FFI\\
2086583008909904896 & 296.9578235 & 47.5964425 & 17.165 & RRC & 10418799 & 17.125 & FFI\\
2100304359969626624 & 285.1271038 & 39.4015144 & 17.685 & RRC & 4345865 & 17.716 & FFI\\
2100508422455676288 & 287.5221537 & 39.5553736 & 18.080 & RRC & 4451334 & 18.263 & FFI\\
2102965800881905792 & 288.3245722 & 43.5415773 & 15.384 & RRC & 7812805 & 15.546 & FFI\\
2102980060173158144 & 287.9872854 & 43.3258509 & 15.749 & RRC & 7672313 & 15.864 & FFI\\
2103421170493346816 & 284.5119233 & 40.3957360 & 16.541 & RRC & 5166889 & 16.784 & FFI\\
2104875824376341888 & 283.7965235 & 42.0339167 & 16.012 & RRC & 6584320 & 16.053 & FFI\\
2105082292044513536 & 283.5884772 & 43.0995012 & 15.724 & RRC & 7422845 & 15.867 & FFI\\
2105292058247822976 & 281.8111161 & 44.1887140 & 17.201 & RRC & 8211945 & 17.258 & FFI\\
2105292195685604608 & 283.0499329 & 43.4426085 & 17.006 & RRC & 7733600 & 17.038 & FFI\\
2105850953752583680 & 284.8012931 & 43.9498940 & 16.619 & RRC & 8081725 & 16.927 & FFI\\
2106317524639255680 & 286.2291335 & 44.7836872 & 16.065 & RRC & 8612183 & 16.198 & FFI\\
2106527496999999488 & 285.9604747 & 46.0288732 & 13.296 & RRC & 9453114 & 13.419 & target\\
2106890954312188288 & 283.3226482 & 45.0401760 & 15.321 & RRC & 8801073 & 15.56 & FFI\\
2106998156695528448 & 283.3651904 & 45.5335706 & 14.838 & RRC & 9137819 & 14.991 & target\\
2116749617944811264 & 280.8543785 & 42.5792557 & 18.126 & RRC & 7006857 & 17.999 & FFI\\
2126843276427186432 & 290.0433736 & 43.7155517 & 14.864 & RRC & 7954849 & 14.862 & FFI\\
2132728824731016192 & 288.8330915 & 50.2067867 & 16.846 & RRC & 11909124 & 16.706 & FFI\\
2133142859577849856 & 289.2404017 & 50.8121588 & 16.904 & RRC & 12204812 & 17.119 & FFI\\
\hline 
\end{tabular}
\end{table*}

\begin{table*}
\caption{Stars with uncertain but likely classifications from the original \textit{Kepler} observations.  } 
\label{tab:kuncertxmatch}
\centering 
\begin{tabular}{r c c c c c c c c} 
\hline
\hline 
\noalign{\vskip 3pt}
\textit{Gaia} DR2 source\_id & DR2 RA & DR2 Dec & \textit{G}  & DR2 class & KIC & \textit{Kp}  & data type & \textit{Kepler} class\\
 & (deg) & (deg) & (mag) &  &  & (mag) & &\\
  \hline 
\noalign{\vskip 3pt}
2052863426827875456 & 290.6708966 & 38.8452285 & 17.444 & RRAB & 3744225 & 17.710 & FFI & RRab?\\
2073860422436544000 & 297.4519701 & 40.9893283 & 14.835 & RRAB & 5727406 & 14.972 & FFI & RRab?\\
2099377506028502400 & 288.2585772 & 38.4944088 & 19.527 & RRAB & 3331090 & 19.427 & FFI & RRab?\\
2128491611857019776 & 292.9388450 & 47.8195878 & 18.659 & RRAB & 10602685 & 18.171 & FFI & RRab?\\
2126654744543334784 & 291.5287676 & 45.1965107 & 16.213 & RRAB & 8884795 & 16.261 & FFI & RRc?\\
2077964177781795200 & 295.0209805 & 42.9274675 & 17.784 & RRC & 7373841 & 17.267 & FFI & RRc?\\
2078178681323184768 & 294.4612324 & 43.3732497 & 19.252 & RRC & 7690615 & 19.224 & FFI & RRc?\\
2080190551386659456 & 296.2246715 & 45.7374787 & 17.731 & RRC & 9292598 & 17.933 & FFI & RRc?\\
2080284112954652544 & 297.5274861 & 46.4933085 & 17.164 & RRC & 9725750 & 17.141 & FFI & RRc?\\
2086553047221169408 & 298.1033115 & 48.5651069 & 16.978 & RRC & 11045097 & 16.933 & FFI & RRc?\\
2104482469792786688 & 283.3977429 & 41.7620072 & 18.390 & RRC & 6343434 & 18.275 & FFI & RRc?\\
2105141974910639232 & 283.2231541 & 43.9378973 & 14.954 & RRC & 8078670 & 15.143 & FFI & RRc?\\
2130857971336396032 & 287.5319452 & 47.2799976 & 17.003 & RRC & 10198186 & 17.157 & FFI & RRc?\\
2132959408636529792 & 287.6619814 & 50.4778068 & 19.039 & RRC & 12006372 & 18.943 & FFI & RRc?\\
2052170249169053568 & 294.1764164 & 38.5791661 & 19.825 & RRAB & 3454674 & 18.810 & FFI & RRc/EB?\\
2085218132725526528 & 300.3009309 & 45.8995148 & 19.121 & RRAB & 9368129 & 18.842 & FFI & RRc/EB?\\
2100223412723866880 & 284.7592318 & 39.2885991 & 18.378 & RRC & 4136159 & 18.187 & FFI & RRc/EB?\\
2133079873879600384 & 289.9087494 & 50.5891289 & 17.925 & RRC & 12059064 & 17.620 & FFI & RRc/EB?\\

\hline 
\end{tabular}
\end{table*}

\begin{table*}
\caption{Crossmatch of RRab-type stars in the \textit{K2} observations, \textit{K2} C refers to the observing campaign. The \textit{Kp} brightness refers to the flux-calibrated values. For the \textit{K2} observations we also list the pulsation periods. The full table is available electronically. } 
\label{tab:rrabxmatch}
\centering 
\begin{tabular}{r c c c c c c c c} 
\hline
\hline 
\noalign{\vskip 3pt}
\textit{Gaia} DR2 source\_id & DR2 RA & DR2 Dec & \textit{G} & DR2 Class & EPIC &  \textit{K2} C & \textit{Kp}  & P  \\
 & (deg) & (deg) & (mag) &  & &  & (mag) & (d) \\
\hline 
\noalign{\vskip 3pt}
3374190873985045888 & 94.4518789 & 19.7203845 & 15.109 & RRAB & 202064530 & 0 & 15.159 & 0.49697\\
3376253420360404992 & 97.3380133 & 22.1837461 & 16.122 & RRAB & 202064526 & 0 & 16.044 & 0.56076\\
3379937093550342528 & 101.2078923 & 24.3026555 & 14.439 & RRAB & 202064516 & 0 & 14.531 & 0.52379\\
3381448299265993728 & 101.3520857 & 24.5421433 & 14.789 & RRAB & 202064491 & 0 & 14.565 & 0.57253\\
3425627711558357248 & 93.8482864 & 23.8313169 & 15.895 & RRAB & 202064531 & 0 & 15.409 & 0.50795\\
3602396707054332160 & 179.8946495 & -1.8598515 & 19.006 & RRAB & 201339783 & 1 & 19.573 & 0.53086\\
3794039453472320256 & 174.1404128 & -1.3380417 & 18.571 & RRAB & 201375063 & 1 & 18.591 & 0.62803\\
3795681505368933760 & 176.5094136 & 0.3493481 & 19.828 & RRAB & 201488452 & 1 & 19.561 & 0.59032\\
3796741163995579136 & 172.7052697 & -0.9883901 & 19.391 & RRAB & 201398311 & 1 & 19.624 & 0.46457\\
3797093488752748032 & 171.4873933 & -0.1616250 & 14.282 & RRAB & 201454019 & 1 & 14.326 & 0.70842\\
\multicolumn{9}{l}{$\ldots$}\\
\hline 
\end{tabular}
\end{table*}

\begin{table*}
\caption{Crossmatch of RRc-type stars in the \textit{K2} observations. The full table is available electronically. } 
\label{tab:rrcxmatch}
\centering 
\begin{tabular}{r c c c c c c c c} 
\hline
\hline 
\noalign{\vskip 3pt}
\textit{Gaia} DR2 source\_id & DR2 RA & DR2 Dec & \textit{G} & DR2 Class & EPIC &  \textit{K2} C & \textit{Kp}  & P  \\
 & (deg) & (deg) & (mag) &  & &  & (mag) & (d) \\
\hline 
\noalign{\vskip 3pt}
3786001409293244800 & 173.4343361 & -5.1026035 & 15.862 & RRC & 201158092 & 1 & 16.061 & 0.31014\\
3896812458883270400 & 175.4991113 & 3.9916983 & 15.412 & RRC & 201720727 & 1 & 15.477 & 0.30975\\
66183697982876416 & 60.1350351 & 24.4304789 & 16.015 & RRC & 211091644 & 4 & 16.041 & 0.29042\\
607162729019029888 & 135.5588208 & 14.1677961 & 16.358 & RRC & 211573254 & 5 & 16.495 & 0.31959\\
609116694325015680 & 130.7046858 & 13.3769410 & 16.293 & RRC & 211516905 & 5 & 16.275 & 0.34025\\
612376063402765312 & 135.0281818 & 18.6079110 & 15.357 & RRC & 211891936 & 5 & 15.386 & 0.34697\\
657965796923993216 & 129.2065123 & 15.9333259 & 16.456 & RRC & 211701322 & 5 & 16.311 & 0.33911\\
658334137615243648 & 131.1006087 & 17.2093019 & 19.544 & RRC & 211792469 & 5 & 19.968 & 0.30588\\
659746632100473984 & 129.2783707 & 18.9322997 & 15.905 & RRC & 211913888 & 5 & 15.942 & 0.2977\\
665174886647140736 & 130.4816877 & 22.0112793 & 15.745 & RRC & 212099502 & 5 & 15.767 & 0.32141\\
\multicolumn{9}{l}{$\ldots$}\\
\hline 
\end{tabular}
\end{table*}

\begin{table*}
\caption{Crossmatch of RRd-type stars in the \textit{K2} observations. } 
\label{tab:rrdxmatch}
\centering 
\begin{tabular}{r c c c c c c c c c} 
\hline
\hline 
\noalign{\vskip 3pt}
\textit{Gaia} DR2 source\_id & DR2 RA & DR2 Dec & \textit{G} & DR2 Class & EPIC &  \textit{K2} C & \textit{Kp}  & $P_0$ & $P_1$  \\
 & (deg) & (deg) & (mag) &  & &  & (mag) & (d) & (d) \\
\hline 
\noalign{\vskip 3pt}
3796490612783265152 & 176.8340702 & 1.8239436 & 15.839 & RRC & 201585823 & 1 & 15.774 & 0.48260 & 0.35942\\
6248239227324924416 & 239.1581229 & -18.8471338 & 14.745 & RRAB & 205209951 & 2 & 14.757 & 0.47077 & 0.34877\\
51156844364167552 & 58.0512607 & 20.3530932 & 15.504 & RRC & 210831816 & 4 & 15.541 & 0.48878 & 0.36380\\
612194609624700928 & 135.5928672 & 18.5610970 & 18.786 & RRC & 211888680 & 5 & 19.185 & 0.48300 & 0.35938\\
3610631916003219328 & 204.2092377 & -11.7283006 & 15.674 & RRD & 212547473 & 6 & 15.619 & 0.54507 & 0.40643\\
3620942277055055488 & 197.4515385 & -13.8228783 & 16.517 & RRC & 212449019 & 6 & 16.617 & 0.48777 & 0.36339\\
4071397068375975296 & 282.0255672 & -29.2124249 & 18.162 & RRC & 229228184 & 7 & 18.793 & 0.45927 & 0.34111\\
4071405658308934144 & 282.1318806 & -29.0042516 & 18.076 & RRAB & 229228194 & 7 & 19.183 & 0.52268 & 0.38971\\
4071509081124919040 & 281.9373883 & -28.3133952 & 18.531 & RRC & 229228175 & 7 & 18.858 & 0.47000 & 0.34945\\
4072051140361909632 & 282.7750015 & -27.2508593 & 15.832 & RRC & 214147122 & 7 & 15.902 & 0.54103 & 0.40366\\
6761560112114793216 & 282.6946631 & -29.1564157 & 16.901 & RRAB & 213514736 & 7 & 17.208 & 0.50359 & 0.37522\\
2576293393286532224 & 16.9326178 & 5.9752228 & 18.050 & RRC & 229228811 & 8 & 18.651 & 0.50021 & 0.37291\\
2580012972403894528 & 19.7918896 & 9.8452558 & 17.320 & RRAB & 220636134 & 8 & 17.390 & 0.50138 & 0.37374\\
3682596906250805632 & 192.1245496 & -2.3467725 & 18.053 & RRD & 228952519 & 10 & 18.500 & 0.55142 & 0.40453\\
3698207256945765760 & 185.5512423 & 0.0526262 & 20.317 & RRAB & 248369176 & 10 & 20.952 & 0.56839 & 0.42407\\
3699831549153899648 & 185.4119454 & 0.8164297 & 17.024 & RRD & 201519136 & 10 & 17.019 & 0.46348 & 0.34432\\
2413839863087928064 & 349.3069278 & -10.1488848 & 16.844 & RRD & 245974758 & 12 & 17.281 & 0.47527 & 0.35331\\
2438582821787698176 & 351.4208931 & -8.0499976 & 17.044 & RRD & 246058914 & 12 & 17.105 & 0.45295 & 0.33611\\
\noalign{\vskip 3pt}
\hline 
\end{tabular}
\end{table*}

\begin{table*}
\caption{Cepheid crossmatch. The three sections represent the \textit{Kepler} field (top), clear identifications in the \textit{K2} fields (middle), and uncertain identifications (bottom). } 
\label{tab:cepxmatch}
\centering 
\begin{tabular}{r c c c c c c c } 
\hline
\hline 
\noalign{\vskip 3pt}
\textit{Gaia} DR2 source\_id & DR2 RA & DR2 Dec & \textit{G} & DR2 class & KIC/EPIC & Campaign & ID\\
 & (deg) & (deg) & (mag) &  &  &  & \\
 \hline
\noalign{\vskip 3pt}
2078709577944648192 & 297.0644335 & 43.1269176 & 8.916 & CEP & 7548061 & Kepler & V1154 Cyg\\
2101097215232231808 & 290.4127674 & 39.9355875 & 10.396 & T2CEP & 4831185 & Kepler & HP Lyr\\
\hline
\noalign{\vskip 3pt}
3378049163365268608 & 100.7812965 & 20.9391061 & 9.676 & CEP & 202064438 & 0 & AD Gem\\
3423693395726613504 & 90.6524429 & 22.2341468 & 9.549 & CEP & 202064436 & 0 & RZ Gem\\
3425495186047291136 & 92.5806547 & 24.0208511 & 10.485 & CEP & 202064435 & 0 & V371 Gem\\
3425576270732293632 & 93.9995283 & 23.7474871 & 11.668 & T2CEP & 202064439 & 0 & BW Gem\\
6249649277967449216 & 242.8120587 & -16.8611619 & 12.444 & T2CEP & 205546706 & 2 & KT Sco\\
47299585774090112 & 65.0074841 & 17.2793921 & 17.040 & T2CEP & 210622262 & 4 & -- \\
4085507620804499328 & 286.6122728 & -19.6098184 & 12.820 & T2CEP & 217987553 & 7 & V1077 Sgr\\
4085983537561699584 & 282.0408160 & -20.1265928 & 13.173 & T2CEP & 217693968 & 7 & V377 Sgr\\
4087335043492541696 & 286.5130651 & -18.4282476 & 12.323 & T2CEP & 218642654 & 7 & V410 Sgr\\
6869460685678439040 & 293.6444501 & -19.3611183 & 12.956 & ACEP & 218128117 & 7 & ASAS J193435-1921.7\\
4052361842043219328 & 274.9408203 & -27.1592265 & 12.806 & T2CEP & 222668291 & 9 & V1185 Sgr\\
4066429066901946368 & 273.2604032 & -23.1172942 & 6.835 & CEP & 225102663 & 9 & 12 Sgr\\
4093976334264606976 & 273.8594387 & -20.6295526 & 10.131 & CEP & 226412831 & 9 & V1954 Sgr\\
4096140001386430080 & 275.8297937 & -18.5747736 & 9.909 & CEP & 227267697 & 9 & AY Sgr\\
4096341040228858240 & 275.2730788 & -18.4554741 & 9.832 & CEP & 227315843 & 9 & V5567 Sgr\\
4096979650282842112 & 276.1854194 & -16.7971738 & 8.315 & CEP & 227916945 & 9 & XX Sgr\\
4118144527610250880 & 264.9723348 & -20.9930069 & 14.470 & T2CEP & 226238697 & 9 & BLG-T2CEP-27\\
4111834567779557376 & 256.5229101 & -26.5805651 & 6.835 & CEP & 232257232 & 11 & BF Oph\\
3409635486731094400 & 69.3115536 & 18.5430127 & 6.267 & CEP & 247086981 & 13 & SZ Tau\\
3415206707852656384 & 76.3094130 & 21.7635904 & 12.300 & T2CEP & 247445057 & 13 & VZ Tau\\
\hline
\noalign{\vskip 3pt}
3423579012158717184 & 92.1462986 & 22.6172200 & 8.377 & CEP & 202062191 & 0 & SS Gem (T2CEP?)\\
4111218875639075712 & 260.2294555 & -23.4355371 & 13.657 & T2CEP & 235265305 & 11 & -- \\
4111880369315900032 & 255.2860252 & -26.5951004 & 11.628 & T2CEP & 232254012 & 11 & ET Oph\\
4112437031430794880 & 257.1786569 & -25.1637301 & 12.962 & T2CEP & 231047453 & 11 & IO Oph\\
2638680812622984960 & 348.8605810 & -1.3746326 & 17.749 & ACEP & 246385425 & 12 & --\\
\noalign{\vskip 3pt}
\hline 
\end{tabular}
\end{table*}

\begin{table*}
\caption{Crossmatch of stars that turned out not to be RR~Lyrae stars. The last column provides alternative identifications and/or likely variability class, if possible.  } 
\label{tab:nonrrmatch}
\centering 
\begin{tabular}{r c c c c c c c } 
\hline
\hline 
\noalign{\vskip 3pt}
\textit{Gaia} DR2 source\_id & DR2 RA & DR2 Dec & \textit{G} & DR2 Class & EPIC &  \textit{K2} C &  Comment  \\
 & (deg) & (deg) & (mag) &  & &   & \\
\hline 
\noalign{\vskip 3pt}
3804350643452878848 & 166.9247344 & 0.4356445 & 18.594 & RRAB & 201494217 & 1 & --\\
2601409430025491200 & 338.7320094 & -13.0709211 & 17.999 & RRAB & 212235345 & 3 & ROT\\
66847287606832768 & 56.7996921 & 25.1162548 & 11.421 & RRAB & 211132787 & 4 & EC\\
4071401500782333696 & 281.9615772 & -29.1119425 & 18.155 & RRAB & 213528187 & 7 & ROT/DSCT\\
4072113675078719360 & 282.0081017 & -27.5777297 & 15.741 & RRAB & 214027794 & 7 & low amplitude\\
4072275715608921344 & 282.9797618 & -26.8576286 & 18.365 & RRAB & 229228258 & 7 & ROT\\
4073145880282355200 & 282.3548453 & -25.8753240 & 20.193 & RRAB & 214689700 & 7 & low amplitude\\
4074665130492370560 & 282.6322558 & -25.3923767 & 16.858 & RRC & 214895832 & 7 & HADS\\
4075390224042082688 & 284.9051268 & -23.3645490 & 14.838 & RRAB & 215881928 & 7 & BL Her (V839 Sgr)\\
4078049770897097088 & 281.5874740 & -24.1344566 & 18.138 & RRAB & 215479005 & 7 & ROT\\
4078775551632704896 & 283.7273346 & -21.9521609 & 15.731 & RRAB & 216660299 & 7 & --\\
4082823506754098432 & 289.0457792 & -20.9322003 & 14.921 & RRAB & 217235287 & 7 & BL Her (V527 Sgr)\\
3698191112165374976 & 184.8087296 & -0.1380778 & 15.335 & RRAB & 201455676 & 10 & instrumental\\
4058445066313103744 & 262.9527077 & -30.1317612 & 18.940 & RRAB & 242184466 & 11 & --\\
4059999702649953152 & 264.3055426 & -29.8639067 & 19.727 & RRAB & 240291558 & 11 & --\\
4060392954243527296 & 263.8477795 & -28.3171369 & 17.458 & RRAB & 240709679 & 11 & ROT\\
4117562851501910912 & 263.5796074 & -22.4795227 & 20.256 & RRAB & 225461305 & 11 & long period\\
4127629876912023168 & 255.7664846 & -21.4563876 & 15.481 & RRAB & 230545230 & 11 & HADS\\
4134649262221615104 & 258.3963306 & -18.0976697 & 14.366 & RRAB & 234649037 & 11 & BL Her (V1637 Oph)\\
4134727499321491072 & 258.7677959 & -18.1450438 & 19.791 & RRAB & 234640705 & 11 & --\\
6030292619420564352 & 255.8996713 & -27.6952328 & 13.949 & RRAB & 251248334 & 11 & BL Her (FZ Oph)\\
2435741447518507392 & 354.9756087 & -9.0838411 & 15.281 & RRAB & 246015642 & 12 & ACEP \\
2636644688886741248 & 345.5868479 & -3.3788560 & 19.691 & RRAB & 246284344 & 12 & flare\\
3411660546628972928 & 74.1444421 & 20.6563863 & 14.468 & RRAB & 247311936 & 13 & long period\\
\noalign{\vskip 3pt}
\hline 
\end{tabular}
\end{table*}

\begin{table*}
\caption{Crossmatch of stars that turned out not to be Cepheid variables. The last column provides alternative identifications and/or likely variability class, if possible.  } 
\label{tab:noncepmatch}
\centering 
\begin{tabular}{r c c c c c c c } 
\hline
\hline 
\noalign{\vskip 3pt}
\textit{Gaia} DR2 source\_id & DR2 RA & DR2 Dec & \textit{G} & DR2 Class & EPIC &  \textit{K2} C &  Comment  \\
 & (deg) & (deg) & (mag) &  & &   & \\
\hline 
\noalign{\vskip 3pt}
2101228774371560832 & 288.8005648 & 39.7139988 & 11.473 & T2CEP & 4644922 & Kepler & SRV (V677 Lyr)\\
\hline 
\noalign{\vskip 3pt}
3799150778087557888 & 173.5765613 & 1.2033677 & 12.936 & T2CEP & 201544345 & 1 & LPV?\\
2594314457585274240 & 337.8754104 & -18.0702930 & 13.192 & T2CEP & 205903217 & 3 & LPV (BC Aqr)\\
2609074232957002880 & 338.0786514 & -9.5948047 & 16.970 & ACEP & 206210264 & 3 & RRab\\
2612858782044502272 & 334.2271302 & -11.8676581 & 13.553 & T2CEP & 206109248 & 3 & --\\
3609358780321640576 & 198.9566943 & -13.7125905 & 15.246 & T2CEP & 212454161 & 6 & EB\\
4087799999464896512 & 288.8008055 & -17.0581624 & 13.837 & T2CEP & 219308521 & 7 & --\\
6762146937758096640 & 284.0731029 & -27.5238093 & 10.667 & CEP & 214047277 & 7 & SRV? (V4061 Sgr)\\
3578288819399570816 & 189.6473674 & -10.9422293 & 11.478 & T2CEP & 228708336 & 10 & LPV?\\
3699052820043280640 & 181.8713363 & 0.6166174 & 13.984 & T2CEP & 201506181 & 10 & --\\
4108800671623776128 & 256.5415354 & -27.1369675 & 10.173 & CEP & 232135078 & 11 & LPV?\\
4116400736547210496 & 264.3714041 & -23.8031748 & 14.632 & T2CEP & 224691021 & 11 & --\\
3392988846324877952 & 75.3718178 & 15.0239664 & 13.588 & T2CEP & 246736776 & 13 & EB\\
3419364167475320448 & 74.4462919 & 23.5072079 & 16.317 & T2CEP & 247671949 & 13 & ROT?\\
3419422583326134144 & 75.0288648 & 24.1428188 & 14.012 & T2CEP & 247761523 & 13 & EB\\
\noalign{\vskip 3pt}
\hline 
\end{tabular}
\end{table*}

\end{document}